 \documentclass[11pt,a4paper]{article}
\usepackage{jheppub}
\def\beq{\begin{equation}}
\def\eeq{\end{equation}}
\def\bea{\begin{eqnarray}}
\def\eea{\end{eqnarray}}

\usepackage{amsmath}
\usepackage{epsfig,epsf}
\usepackage{verbatim}
\usepackage{epstopdf}
\def\eq#1{{Eq.~(\ref{#1})}}
\def\fig#1{{Fig.~\ref{#1}}}
\newcommand{\Lb}{\left(}
\newcommand{\Rb}{\right)}

\title{Reggeon Field Theory for Large Pomeron Loops.}
\author[a]{Tolga Altinoluk,}
\author[b]{Alex Kovner,}
\author[c,d]{Eugene Levin,}
\author[e]{ Michael Lublinsky}

\affiliation[a]{
Departamento de F\'\i sica de Part\'\i culas and IGFAE, Universidade de Santiago de Compostela, E-15782 Santiago de Compostela, Galicia-Spain}
\affiliation[b]{Physics Department, University of Connecticut, 2152 Hillside road, Storrs, CT 06269, USA}
\affiliation[c] {Departamento de F\'\i sica, Universidad T\'ecnica
Federico Santa Mar\'\i a,
and
Centro Cient\'\i fico-Tecnol$\acute{o}$gico de Valpara\'\i so, Avda. Espa\~na 1680,
Casilla 110-V,  Valpara\'\i so, Chile}
\affiliation[d]{Department of Particle Physics,  Tel Aviv University , Tel Aviv 69978, Israel}
\affiliation[e]{Physics Department, Ben-Gurion University of the Negev, Beer Sheva 84105, Israel}
 
\date{\today}
\abstract{We analyze the range of applicability of the high energy Reggeon Field Theory $H_{RFT}$ derived in \cite{aklp}. We show that this theory is valid as long as 
{\it at any intermediate value of rapidity $\eta$} throughout the evolution {\it at least one} of the colliding objects is dilute. Importantly, at some values of $\eta$ the dilute object could be the projectile, while at others it could be the target, so that $H_{RFT}$ does not reduce to either $H_{JIMWLK}$ or $H_{KLWMIJ}$.  When both objects are dense, corrections to the evolution not accounted for in \cite{aklp} become important. 
The same limitation applies to other approaches to high energy evolution available today, such as for example \cite{raju} and \cite{braun}.
We also show that, in its regime of applicability  $H_{RFT}$ can be simplified. We derive the simpler version of $H_{RFT}$ and in the large $N_c$ limit rewrite it in terms of the Reggeon creation and annihilation operators. The resulting $H_{RFT}$ is explicitly self dual and provides the generalization of the Pomeron calculus developed in \cite{braun} by including higher Reggeons in the evolution. It is applicable for description of `large' Pomeron loops, namely Reggeon graphs where all the splittings occur close in rapidity to one dilute object (projectile), while all the merging close to the other one (target). Additionally we derive, in the same regime expressions for single and double inclusive gluon production (where the gluons are not separated by a large rapidity interval) in terms of the Reggeon degrees of freedom.
 
}

\date{\today}
\begin{document}
\maketitle

\pagestyle{empty}
\newpage

\mbox{}

\pagestyle{plain}

\setcounter{page}{1}
\section{Introduction}
This paper addresses the question of formulating the QCD Reggeon Field Theory, which consistently includes Pomeron loop effects. The effects of Pomeron loops become important when the hadronic wave function is evolved to high energy and the saturation physics takes over from the linear BFKL evolution. 

Some years ago three of us have given a QCD derivation of the Hamiltonian of the Reggeon Field Theory $H_{RFT}$ \cite{aklp}. The derivation of \cite{aklp}, accounted for both 
effects that are important in the nonlinear high energy regime - the nonlinear corrections to the evolution of the wave function of a dense projectile \cite{KLUrs}, as well as multiple scattering corrections important for dense target. The derivation used perturbation theory in strong external field in order to calculate the soft photon wave function of the dense projectile. 

However it was subsequently realized that this approximation is inadequate to describe scattering of a dense projectile on a dense target. For such a process one needs to know the large field ``tail'' of the wave function, namely the part of the wave function that carries very little probability density, but nevertheless determines the overlap between the incoming and outgoing projectile states, when such a state is strongly altered in the scattering process. This part of the problem cannot be addressed in perturbation theory in external field, but requires a more elaborate semiclassical treatment. 

Nevertheless, the $H_{RFT}$ derived in \cite{aklp} reduces to the two known limits - JIMWLK and KLWMIJ - in the approximation of a dilute target or dilute projectile respectively. We have argued that it adequately  takes into account the Pomeron loops in the situation of scattering of two dilute objects at very high energy. 

In this paper we analyze more carefully in what circumstances $H_{RFT}$ is applicable. We show that it is valid as long as 
at any intermediate value of rapidity $\eta$ throughout the evolution {\it at least one} of the colliding objects is dilute. We also show that in this regime $H_{RFT}$ can be simplified. We derive the simpler version of the Hamiltonian and rewrite it explicitly in terms of QCD Reggeon operators and their duals (conjugates) discussed previously in \cite{last}. We also consider gluon production in this regime and derive compact self dual expressions for the single and double inclusive gluon production cross sections.

In Section 2 we review the basic setup of our approach. In Section 3 we express $H_RFT$ in
terms of color singlet Reggeon operators and discuss its properties. Section 4 reformulates
gluon production amplitudes in terms of the Reggeons introduced in Section 3. We present
our conclusions in Section 5. Several Appendices supplement calculations of Sections 3 and 4.

\section{The High Energy Evolution.}
Our main object of interest in the first part of this paper is the evolution of a scattering matrix of two hadrons, which we refer to as ``projectile'' and ``target''. In the high energy eikonal approximation the scattering matrix is calculated as  
\begin{equation}\label{s}
\langle {\cal S}\rangle=\int d\alpha\int d\rho\delta(\rho)W_P[\delta/\delta\rho]\,e^{ig^2\int_x\rho^a(x)\alpha^a(x)}W_T[\alpha]
\end{equation}
The simple exponential form of the $S$-matrix operator $e^{ig^2\int_x\rho^a(x)\alpha^a(x)}$ 
is the result of the eikonal approximation, while the functional integrals over $\rho(x)$ and $\alpha(x)$ represent averaging over the projectile and target wave functions respectively. Here $x$ is the transverse coordinate.
In this expression $\rho^a(x)$ refers to the color charge density of the projectile, while $\alpha^a(x)$ to the color field of the target, so that strictly speaking we should write 
$\rho_P^a(x)$ and $\alpha_T^a(x)$. We have dropped the subscripts $P$ and $T$ for simplicity, and will follow this practice in the rest of this paper, unless we need to explicitly differentiate between the projectile and target degrees of freedom.
We have written the projectile probability density function in terms of its functional Fourier transform $W_P[\delta/\delta\rho]$ for future convenience. The target probability density can also be written in this form, but we choose to keep it as a function of the color  $\alpha$.

The natural way to write the functional $W_P$ in the dilute projectile limit is as a function of the unitary matrix 
\begin{equation}\label{R}
R(x)=e^{T^a\delta/\delta\rho^a(x)}
\end{equation}
where $T^a$ is the generator of the $SU(N_c)$ group in the fundamental representation.
When acting on the exponential eikonal $S$-matrix, every power of $\delta/\delta\rho$ turns into $i\alpha$, and thus $R(x)\rightarrow \exp\{ig^2T^a\alpha^a(x)\}$, which is just the eikonal phase factor for propagation of a projectile parton at transverse coordinate $x$ through the target field $\alpha(x)$.

For a fixed configuration of $n$ partons at transverse coordinates $x_1,...x_n$ the functional $W_P$ has the form
\begin{equation}
W_P=R(x_1)R(x_2)...R(x_n)
\end{equation}
which upon integration over $\rho$ in eq.(\ref{s}) gives the eikonal phase factor for the state of $n$ partons propagating through the target field.
In the following we will be interested in the propagation of color neutral states, which implies that the left and right indices of $R(x_i)$ are all contracted into color singlets. 
A particular example is  a projectile consisting of a single color dipole. The functional $W_P$ of a single dipole is given by
\begin{equation}
W_P^{dipole}=\frac{1}{N_c}tr[R^\dagger(x)R(y)]
\label{dipoleamp}
\end{equation}
As the energy of the process is increased, the evolution of the $S$-matrix is given by the action of the Reggeon Field Theory Hamiltonian $H_{RFT}$. 
\begin{equation}\label{evol}
\frac{d}{d Y}\langle {\cal S}\rangle=\int d\alpha\int d\rho\delta(\rho)W_P[\delta/\delta\rho]H_{RFT}[\rho,\delta/\delta\rho]\,e^{ig^2\int_x\rho^a(x)\alpha^a(x)}W_T[\alpha]
\end{equation}

Two particular limits of $H_{RFT}$ have been known for a while now. When the projectile is dilute (the color charge density is small), while the target is dense  
the appropriate limit is the KLWMIJ Hamiltonian \cite{klwmij}

\begin{equation}
H_{KLWMIJ}=\frac{\alpha_s}{2\pi^2}\int_{x,y,z}{K_{xyz}\left\{J^a_L(x)J^a_L(y)+J^a_R(x)J^a_R(y)-2J^a_L(x)R^{ab}(z)J^b_R(y)\right\}}
\label{klwmij}
\end{equation}
with the kernel
\begin{equation}
K_{x,y;z}=\frac{(x-z)_i(y-z)_i}{(x-z)^2(y-z)^2}
\end{equation}
The left and right rotation generators when acting on functions of $R$ have the representation
\begin{eqnarray}\label{LR}
J^a_L(x)=tr\left[\frac{\delta}{\delta R^{T}_x}T^aR_x\right]-tr\left[\frac{\delta}{\delta R^{*}_x}R^\dagger_xT^a\right]  \\  
J^a_R(x)=tr\left[\frac{\delta}{\delta R^{T}_x}R_xT^a\right] -tr\left[\frac{\delta}{\delta R^{*}_x}T^aR^\dagger_x\right]
\end{eqnarray}
When $H_{KLWMIJ}$ acts on gauge invariant operators (operators invariant under $SU_L(N_c)$ and $SU_R(N_c)$ rotations of $R$),
 the kernel $K_{xyz}$ can be substituted by the so called dipole kernel
\begin{equation}
K_{x,y;z}\rightarrow -\frac{1}{2}M_{x,y;z}; \ \ \ \ \ \ \ \ M_{xy;z}=\frac{(x-y)^2}{(x-z)^2(y-z)^2}
\end{equation}
The $SU(N_c)$ rotation generators can be expressed explicitly in terms of the color charge density $\rho$ as \cite{last}
\begin{equation}\label{js}
J^a_L(x)=\rho^b(x)\left[\frac{\tau(x)}{2}\coth\frac{\tau(x)}{2}-\frac{\tau(x)}{2}\right]^{ba};\ \ \ \ 
J^a_R(x)=\rho^b(x)\left[\frac{\tau(x)}{2}\coth\frac{\tau(x)}{2}+\frac{\tau(x)}{2}\right]^{ba}
\end{equation}
where
\begin{equation}
\tau(x)\equiv t^a\frac{\delta}{\delta\rho^a(x)}
\end{equation}
with $t^a_{bc}= if^{abc}$ - the generator of $SU(N_c)$ in the adjoint representation.

This is the regime where the evolution is dominated by the so called Pomeron splittings. The evolution of the wave function of the dilute projectile is dominated 
by emission of soft gluons. These gluons proliferate exponentially, and multiple Pomeron exchanges have to be taken into account.

In the opposite regime, where the projectile is dense, but the target is dilute the relevant evolution is given by JIMWLK Hamiltonian\cite{jimwlk}. It is obtained from $H_{KLWMIJ}$ by the dense-dilute duality (DDD) transformation \cite{duality,dipoles}
\begin{equation}
R(x)\rightarrow S^F(x)\equiv e^{ig^2T^a\alpha^a_P(x)}
\end{equation}
where $\alpha_P$ is the projectile color field related nonlinearly to the projectile color charge density by
\begin{equation}\label{srho}
\frac{i}{g^2}\partial_i\left[S^{F\,\dagger}(x)\partial_iS^F(x)\right]=T^a\rho_P^a(x)
\end{equation}
DDD transforms the right (left) rotation operators acting on $R$ into right (left) rotation operators acting on $S$.

This regime is dominated by  "Pomeron mergings". For a dense projectile wave function the rate of emission of gluons in the wave function is slowed down 
by the effects of coherent radiation from multiple sources.

While the KLWMIJ and JIMWLK regimes are applicable in the situation where one of the colliding objects is dense and the other one is dilute, 
one is in general interested also in a different situation, where the same object can evolve from dilute to dense during the rapidity evolution.

 In this case one has to include in the evolution the so called Pomeron loops, since the initial stages of the evolution
of a dilute projectile are dominated by Pomeron splittings, while in the final stages Pomeron mergings take over, 
thus creating "large" Pomeron loops. 
Another related issue is that, since the color charge density is dimensionfull, a statement like "large color density" does not make sense by itself. 
Whether the charge color density is large or small, depends on the resolution scale on which one is measuring it. It is more appropriate to think
 of the color charge density as defining an intrinsic scale - the saturations scale $Q_s$ in the hadronic wave function. 
 For momenta below the saturation momentum the wave function looks dense and effects of coherence are important, 
 while for momenta above $Q_s$ the hadron looks dilute. Thus depending on the observable of interest, the same object should be treated 
either as dilute or as dense, which again requires taking account of Pomeron loops.

Several groups have approached the problem of Pomeron loops in the past several years with varying degree of rigor \cite{braun,msw,iancu,balit1,hatta,levinlublinsky,aklp,levin,Avsar}.  In this paper we continue the approach of \cite{aklp}.

Ref.\cite{aklp} provided the most detailed derivation of the high energy evolution, starting from the fundamental QCD Hamiltonian, and considering
 the construction of the soft gluon wave function. This wave function was calculated perturbatively in the coupling constant, but allowing for the 
 presence of large valence color charge density as a source for soft gluons. This wave function then was allowed to scatter eikonally on a dense target, 
 and the resulting scattering matrix determined the evolution Hamiltonian $H_{RFT}$. The resulting Hamiltonian is\footnote{It is worthwhile mentioning that $H_{RFT}$ describes interactions of Reggeons as shown in Ref.\cite{aklp}, and corresponds to Reggeon Field Theory (RFT). This in spite of the fact that it is not written explicitly in terms of Reggeon degrees of freedom. The simplified form of this Hamiltonian that describes the BFKL Pomerons and their interaction in terms of the Pomeron fields we will refer to as the BFKL Pomeron calculus (see Ref.\cite{braun} and eq.(\ref{simple})).}:
\begin{eqnarray}\label{hg1}
H_{RFT}&=&
\frac{1}{8\pi^3 }\,\int_{x,y,z,\bar z}[b_{Ri}^b(x)\,R^{\dagger ba}(x)\,-\,b^a_{Li}(x)]\ \left[\delta_{ij}\frac{1}{(x-z)^2}\,-\,
2\,\frac{(x-z)_i\,(x-z)_j}{ (x-z)^4}\right]\nonumber \\ \nonumber \\ 
&\times&\left[\delta^{ac}\,+\, [S_L^{A\,\dagger}(x)\,S_L^A(z)]^{ac}\right]
\  \tilde K^{-1\,cd}_{\perp jk}(z,\bar z)\ \left[\delta_{kl}\frac{1}{(y-\bar z)^2}\,-\,2\,\frac{(y-\bar z)_k\,(y-\bar z)_l}{(y-\bar z)^4}\right]
\nonumber \\ \nonumber \\ 
&\times&\left[\delta^{de}\,+\, [S_R^{A\,\dagger}(\bar z)\,S_R^A(y)]^{de}\right]\ [b_{Rl}^e(y)\,-\,R^{\dagger ef}(y)\,b^f_{Lk}(y)]
\end{eqnarray}
with
\begin{eqnarray}
\tilde K^{\ ab}_{\perp ij}(x,y)&=&\frac{1}{2\,\pi^2}\,\int_z\left[\delta_{ik}\frac{1}{(x- z)^2}\,-\,2\,\frac{(x- z)_i\,(x- z)_k}{(x- z)^4}\right]
\ \left[\delta_{kj}\frac{1}{(z-y)^2}\,-\,2\,\frac{(z-y)_k\,(z-y)_j}{(z-y)^4}\right]\nonumber\\
&\times&\left\{R^{\dagger ab}(z)\,+\,\left[S_R^{A\,\dagger}(x)\,S_R^A(z)\,R^{\dagger }(z)\,S^{A\,\dagger}_L(z)\,S_L^A(y)\right]^{ab}\right\}
\end{eqnarray}
and
\begin{equation}
b^a_{L(R)i}\,=\,-\frac{1}{g}\,f^{abc}(S_{L(R)}^{A})^{\dagger bd}\,\partial_i \,S_{L(R)}^{A\,dc}\,.
\end{equation}
Here the adjoint unitary matrix $S_{L(R)}^A$ is the adjoint representation of the fundamental matrix $S^F_{L(R)}$ defined as in eq.(\ref{srho}) with 
$J_L (J_R)$ on the right hand side.

As shown in \cite{aklp} this Hamiltonian reduces to the correct JIMWLK and KLWMIJ forms in the appropriate limits. Thus, we have here a Hamiltonian, derived directly from QCD which is applicable in both, dense and dilute limits. Clearly therefore, when applied to evolution of a dilute projectile to very high rapidity, it does contain Pomeron loop contribution. The limits of applicability of this Hamiltonian however, have not been clarified. This is our aim in the next section.

\section{Applicability of $H_{RFT}$.}

In the derivation of $H_{RFT}$ in \cite{aklp} the projectile wave function was treated as having an arbitrary density of color charge. More precisely the density parametrically cannot exceed $1/\alpha_s$ but is allowed to be of order $1/\alpha_s$. The target fields were also assumed to be at most of the same order, and thus the scattering matrix $R$ was assumed to be $O(1)$. Nevertheless, $H_{RFT}$ as derived in \cite{aklp} can not be used for processes involving the scattering of a dense projectile on a dense target. 

We now explain the reason for that. The procedure of calculating $H_{RFT}$ adopted in \cite{aklp,KLUrs} is the following. One first calculates the wave function of the soft gluons of the projectile in the presence of a (possibly large) valence color charge density $\rho^a(x)$, $\Psi_{in}[A,\rho]$. The fields $A$ are soft gluon fields which occupy a small rapidity interval $\delta Y$. This hadronic wave function after scattering on the target becomes $\Psi_{out}[A,\rho]=\Psi_{in}[RA, R\rho]$. The Hamiltonian is obtained by calculating the scattering matrix and expanding it to linear order in $\delta Y$,
\begin{equation}\label{overlap}
\int DA \Psi^*_{in}[A,\rho]\Psi_{in}[RA, R\rho]=1-H_{RFT}[\rho, R]\delta Y+...
\end{equation}

In\cite{aklp,KLUrs} the wave function $\Psi_{in}$ was calculated perturbatively in $\alpha_s$, re-summing to all orders terms of the type $(\alpha_s\rho)^n$. The resulting wave function turns out unsurprisingly Gaussian, and schematically has the form
\begin{equation}\label{psiin}
\Psi_{in}\,=\,{\cal N}\, e^{b[\rho]A-\frac{1}{2}A\Lambda[\rho] A}
\end{equation}
where the classical field $b[\rho]$ at large $\rho$ is parametrically $b\sim O(1/g)$, while the width of the Gaussian $\Lambda \sim O(1)$. 

As in any perturbative calculation, the perturbative expression for $\Psi_{in}$ gives good approximation for the wave function in the region of field space that contains most of the probability density, that is for those values of the field where $\Psi$ is not exponentially small. In the present case this means for $A=\Lambda^{-1}b\pm \Delta$ with $\Delta \sim O(1)$. On the other hand, in the outgoing wave function $\Psi_{out}$, the maximum of the distribution is at a different value of the field $A=\bar \Lambda^{-1}\bar b$ with $\bar b=Rb[R\rho]$ and $\bar \Lambda\sim O(1)$ as before. Thus the overlap of the two wave functions in eq.(\ref{overlap}) is in fact dominated by the tails of the two probability distributions, where the values of the field are far from the maximum of either wave function by the amount of order $1/g$. These tails are not correctly given by the perturbative calculation, and thus the situation where a dense projectile scatters on a dense target is outside the validity of \cite{aklp}. 

On the other hand when either one of the scattering objects is small, it turns out that $\bar b-b\sim O(g)$, and thus $H_{RFT}$ derived in  \cite{aklp} is applicable.
As we demonstrate below the regime of applicability of $H_{RFT}$ extends, in fact to a wider range of situations.

\subsection{Dipole-dipole scattering.}
Let us consider scattering of two dilute systems, which we will figuratively refer to as dipoles, at some high energy. We attribute all the evolution to the wave function of one of the dipoles, "the projectile". Formally, the S-matrix is given by the following expression
\begin{equation}\label{evoldip}
\langle {\cal S}_Y\rangle=\int d\alpha\int d\rho\delta(\rho)W_P[R]e^{-H_{RFT}[\rho,R]Y}\,e^{ig^2\int_x\rho^a(x)\alpha^a(x)}W_T[\alpha]
\end{equation}
Since both, the projectile and the target are initially dilute, not all the terms in $H_{RFT}$ are equally important. To understand this, let us divide the exponential in eq.(\ref{evoldip}) into product of three factors
\begin{equation}\label{exp}
e^{-H_{RFT}[\rho,R]Y}=e^{-H_{RFT}[\rho,R]\eta}e^{-H_{RFT}[\rho,R](Y-\eta-\eta')}e^{-H_{RFT}[\rho,R]\eta'}
\end{equation}
The rightmost factor can be thought of as acting on the target wave function. Technically, since $H_{RFT}$ is a Hermitian operator, we can take it to act on the eikonal factor in 
eq. ({\ref{evoldip}). Then every factor of $R$ in $H_{RFT}$ becomes the matrix $S$ written in terms of the target field $\alpha$, whereas every factor of $\rho$ becomes $\delta/\delta\alpha$. In other words $H_{RFT}$ acting on $e^{ig^2\int \rho\alpha}$ is equivalent to the dual to $H_{RFT}$ written in terms of the target degrees of freedom, acting on $W_T[\alpha]$. As shown in \cite{duality}, the complete $H_{RFT}$ has to be self dual. Although it has not been shown explicitly that $H_{RFT}$ derived in \cite{aklp} has this property, as we will see later it is indeed self dual in its region of applicability. Thus the rightmost factor in eq.(\ref{exp}) evolves $W_T$ to rapidity $\eta'$. 

The leftmost factor in eq.(\ref{evoldip}) acts to the left on the projectile wave function and evolves it to rapidity $\eta$.
Let us take rapidity $\eta$ small enough, so that the evolved projectile wave function
\begin{equation}\label{weta}
W^\eta_P[R]=W_P[R] e^{-H_{RFT}[\rho,R]\eta}
\end{equation}
still describes a dilute system. 

Let us consider the mechanics of the calculation of the evolved probability density eq.(\ref{weta}). To find $W^\eta_P[R]$ one has to commute all the factors of $\rho$ in $ e^{-H_{RFT}}[\rho,R]$ to the left of all the factors of $R$, whereby they vanish upon hitting the $\delta$-function in eq. (\ref{evoldip}). The Hamiltonian $H_{RFT}$ eq.(\ref{hg1}) is a function of $g\rho$ only. Thus any extra power of $\rho$ in the expansion necessarily comes with an extra power of the coupling constant. The smallness of the coupling constant can only be overcome by a large combinatorial factor if $\rho$ has to be commuted through a product of many factors $R$. 

Since initially $W_P[R]$ describes a dilute system, it does not contain many factors of $R$, and therefore $H_{RFT}$ can be safely expanded to lowest order in $\rho$. This is also true for subsequent step in the evolution, as long as $W^\eta_P[R]$ corresponds to a dilute system. As discussed above, to leading order in $\rho$, the Hamiltonian $H_{RFT}$ reduces to the KLWMIJ Hamiltonian  eq. (\ref{klwmij}). The rapidity $\eta$ is of course arbitrary, and can be taken arbitrarily large as long as the condition of diluteness of the projectile evolved to rapidity $\eta$ is satisfied. Within the usual saturation picture this maximal rapidity is of order
 \begin{equation}
 \eta_{max}=\frac{1}{\omega_P}\, \ln\frac{1}{\alpha^2_s}\,\; \ \ \ \ \ \ \ \ \ \ \ \ \ \ \ \omega_P=\,2\,\ln 2\,\alpha_s\,N_c/\pi
 \end{equation}

 A completely analogous argument on the target side tells us that the target probability density effectively evolves with the KLWMIJ Hamiltonian as well, as long as the  target evolved to rapidity $\eta'$ remains dilute. This means that in the rightmost factor in eq.(\ref{exp}) $H_{RFT}$ can be expanded to leading order in $\delta/\delta\rho$, since expansion in powers of $\delta/\delta\rho_P$ translates into expansion in powers of $\alpha_T$ when acting on $W_T[\alpha]$. To leading order in $\delta/\delta\rho$ the Hamiltonian $H_{RFT}$ is equivalent to the JIMWLK Hamiltonian. This argument again is valid as long the rapidity $\eta'$ is smaller than $\eta_{max}$. 
 
 Since $H_{RFT}$ reproduces both the JIMWLK and KLWMIJ evolution in the appropriate limits, it certainly generates valid evolution in these two parts of the rapidity interval. It follows, that for the total rapidity
 \begin{equation}\label{max}
 Y\,<\,2\,\eta_{max}
 \end{equation}
 the Hamiltonian $H_{RFT}$ of \cite{aklp} generates correct high energy evolution. 
 
  Alternatively one can describe this physical situation in the following way. The evolution can be partitioned between the projectile and the target in an arbitrary way. Say, we choose an arbitrary rapidity $0<\eta<Y$ and evolve the projectile wave function by $\eta$ and the target wave function by $Y-\eta$. $H_{RFT}$ is valid if for any $\eta$ in the interval, at least one of the evolved objects is dilute. If the total energy of the process is too large, at some intermediate rapidity one necessarily encounters the situation when both colliding objects are dense. At this energy the perturbative approximation made in deriving $H_{RFT}$ breaks down, and $H_{RFT}$ is not applicable. Essentially, $H_{RFT}$ gives consistent Hamiltonian representation of evolution in the regime discussed in \cite{muellersalam}. 
    
 The estimate of maximal rapidity eq.(\ref{max}) assumed that initially both, the target and the projectile are dilute (dipoles). If this is not the case the argument above restricts the total allowed rapidity $Y$ depending on the initial projectile and target states. Interestingly, this demonstrates the limitation of JIMWLK evolution even in the dense regime. Suppose we consider the scattering of a dense object on a dilute one. Initially the evolution of the dense object is governed by 
  $H_{JIMWLK}$. However this is only correct for rapidities for which the target remains dilute if evolved all the way to the rapidity of the projectile, namely for 
  $Y-Y_0<\eta_{max}$. Once the total evolution rapidity exceeds this value, $H_{JIMWLK}$ has to be amended to take into account scattering of two dense objects.  Thus JIMWLK (and KLWMIJ) evolution can only be consistently applied only if at initial rapidity $Y_0$ the projectile is dense and the target is dilute, and then only up to rapidity $Y=Y_0+\eta_{max}$. This restriction has to be kept in mind while using the JIMWLK equation.

\subsection{Simplifying $H_{RFT}$.}
The preceding discussion does not only establish the validity of $H_{RFT}$ in the parametric regime of eq.(\ref{max}), but also suggests that in this regime its form can be significantly simplified. Consider the following approximation for $H_{RFT}$
\begin{equation}\label{approximation}
H_{RFT}\approx H_{KLWMIJ}+H_{JIMWLK}-H_{BFKL}
\end{equation}
As we have discussed above, in the leftmost exponential factor in eq. (\ref{exp}}) where $H_{RFT}$ acts on dilute projectile, it reduces to $H_{KLWMIJ}$, since one is allowed to expand to lowest order in $\rho$. On the other hand, to lowest order in $\rho$ we know that $H_{JIMWLK}$ reduces to $H_{BFKL}$. Thus eq.(\ref{approximation}) indeed reproduces the correct behavior. Analogously in the exponential terms close to the dilute target expansion in $\delta/\delta\rho$ is valid, and thus $H_{RFT}$ reduces to 
$H_{JIMWLK}$. In the leading order in $\delta/\delta\rho$, $H_{KLWMIJ}$ reduces to $H_{BFKL}$ and thus again eq.(\ref{approximation}) is appropriate. Note that as long as we are at low enough rapidity defined by eq.(\ref{max}), there is always an interval within the evolution, corresponding to the middle exponential factor in eq.(\ref{exp}) where both objects are dilute. In this regime simultaneous expansion in powers of $\rho$ and $\delta/\delta\rho$ is valid and the evolution reduces to the linear BFKL evolution. Again, in this regime eq.(\ref{approximation}) has the correct behavior.
Thus we conclude, that eq.(\ref{approximation}) is parametrically correct approximation  in the range of validity of $H_{RFT}$.

Although this is certainly a simplification, the Hamiltonian eq.(\ref{approximation}) is still fairly complicated. The complexity stems from the fact that the commutation relations between the basic variables of $H_{KLWMIJ}$ and $H_{JIMWLK}$ are in principle very complicated. Recall, that the basic field theoretical variable of the KLWMIJ theory is the unitary matrix $R$ defined in eq.(\ref{R}), whereas the basic variable of JIMWLK is the unitary matrix $S$, which is a complicated nonlinear function of the color charge density determined through the solution of eq.(\ref{srho}). For small $\rho$ it is easy to find $S$, which in the adjoint representation is
\begin{equation}
S^{A\,ab}(x)\approx \delta^{ab}-ig^2t^{ab}_c\frac{1}{\nabla^2}(x,y)\rho^c(y)
\end{equation}
with $t^{ab}_c=if^{abc}$, the generator of $SU(N_c)$ group in the adjoint representation. However for large $\rho$ no explicit expression is available, and consequently the commutation relation between $R$ and $S$ is not known. However, as we demonstrate now, in the regime discussed here one can approximate the commutation
 relation by a simple one.

For simplicity in the rest of this paper we work in the large $N_c$ limit. In this limit, as was shown in \cite{last} the KLWMIJ Hamiltonian can be expressed in terms of 
Reggeon operators and their conjugates. The simplest such Reggeon is the Pomeron, defined in \cite{last} as
\begin{equation}
P(x,y)=1-\frac{1}{2N_c}{\rm Tr}[ R^\dagger (x)R(y)+R^\dagger(y)R(x)]
\end{equation}
We will disregard the other Reggeons for now, but will include them in the Hamiltonian later on.
In terms of the Pomeron, the KLWMIJ Hamiltonian can be written as 
\begin{equation}\label{pom}
H_{KLWMIJ} =-\frac{\alpha_s N_c}{2\,\pi^2}\,\int_{x,y,z} M_{x,y;z}\,\Big\{[P_{x,z}+P_{z,y}\,-\,P_{x,y}-P_{x,z}P_{z,y}]\,P^\dagger_{x,y}\Big\}
\end{equation}

The Pomeron conjugate operator is defined by the relation 
\begin{equation}\label{conp}
[P_{xy},P^\dagger_{uv}]=\delta^+[(uv)-(xy)]
\end{equation}
where 
\begin{equation}
\delta^+[(uv)-(xy)]=\frac{1}{2}[\delta^2(x-u)\delta^2(y-v)+\delta^2(x-v)\delta^2(y-u)]
\end{equation}
To be more precise, we do not require the operator commutation relation eq.(\ref{conp}). Rather, in eq.(\ref{conp}) the operator $P^\dagger$ is understood to act to the left on $P$  which has been constructed explicitly as a function of the unitary matrix $R$. This is equivalent to requirement that $P^\dagger$ has the following matrix elements
\begin{equation}\label{matr1}
\int d\rho \delta[\rho]P_{xy}P^\dagger_{uv}\equiv\langle P_{xy}|P^\dagger_{uv}\rangle=\delta^+[(uv)-(xy)]
\end{equation}
while
\begin{equation}\label{matr2}
\int d\rho \delta[\rho]P^\dagger_{uv}P_{xy}\equiv\langle P^\dagger_{uv}| P_{xy}\rangle=0
\end{equation}
Equations (\ref{matr1},\ref{matr2}) are our working definition for the operator $P^\dagger_{xy}$. We will continue using the notation of eq.(\ref{conp}) in the sense of eqs.(\ref{matr1},\ref{matr2}). Similar notations will be adopted for other Reggeons in the rest of this paper. 


 One can find in principle an explicit expression for $P^\dagger$ in terms of the color charge density operators $\rho$ and the Pomeron field $P$. In \cite{last} it was shown that to lowest order in $P$ one has 
\begin{equation}\label{rhorho}
\frac{1}{N_c}\rho^a(x)\rho^a(y)=P^\dagger_{xy}-\delta_{xy}\int_zP^\dagger_{xz}
\end{equation}

Similarly JIMWLK Hamiltonian can be written in terms of dual Reggeon operators. Let us define the ``dual Pomeron''
\begin{equation}\label{dualP}
\bar P(x,y)=1-\frac{1}{2N_c}{\rm Tr}[ S^{F\,\dagger} (x)S^F(y)+S^{F\,\dagger}(y)S^F(x)]
\end{equation}
Then 
\begin{equation}\label{pomd}
H_{JIMWLK} =-\frac{ \alpha_sN_c}{2\,\pi^2}\,\int_{x,y,z} M_{x,y;z}\,\Big\{[\bar P_{x,z}+\bar P_{z,y}\,-\,\bar P_{x,y}-\bar P_{x,z}\bar P_{z,y}]\,\bar P^\dagger_{x,y}\Big\}
\end{equation}
Here, again we have neglected the contributions of the dual Reggeons except for the dual Pomeron, which we will include later. The conjugate dual Pomeron can also be constructed explicitly. To lowest order
\begin{equation}
\frac{1}{g^4N_c}\nabla^2_{x}\frac{\delta}{\delta \rho^a(x)}\nabla^2_{y}\frac{\delta}{\delta \rho^a(y)}=\bar P^\dagger_{xy}-\delta_{xy}\int_z\bar P^\dagger_{xz}
\end{equation}

In the Hilbert space of the Reggeon field theory, $P$ and $\bar P$ are not independent variables. Clearly, $\bar P$ is closely related to $P^\dagger$, since both are expressible in terms of the color charge density operators, however in general the relation is complicated. Nevertheless in the limit of small charge color density the relation is simple
\begin{equation}\label{aprbar}
\bar P(xy)\approx \frac{g^4}{4}[\frac{1}{\nabla^2}(xu)- \frac{1}{\nabla^2}(yu)][\frac{1}{\nabla^2}(xv)- \frac{1}{\nabla^2}(yv)][P^\dagger_{uv}-\delta_{uv}\int_zP^\dagger_{uz}]
\end{equation}
In the limit where eq.(\ref{aprbar}) is valid, the commutator between $P$ and $\bar P$ is simple. In terms of
\begin{equation}
\Delta(x, y; u, v)\equiv\Big(\frac{1}{\nabla^2}( xu)-\frac{1}{\nabla^2}( yu)-\frac{1}{\nabla^2}( xv)
+\frac{1}{\nabla^2}( yv)\Big)
\end{equation}
it becomes
\begin{equation}\label{commut}
[P( x y),\bar P(uv)]=\frac{g^4}{8}\Big[\Delta( x,  y; u, v)\Big]^2
\end{equation}

This is just the scattering amplitude of a dipole with coordinates $(x,y)$ on a dipole with coordinates $(u,v)$, evaluated in the two gluon exchange approximation.
 This expression  is of course very natural. While the meaning of $R(x)$ as we have discussed above, is that of the scattering amplitude of a parton in the projectile wave function, the meaning of $S(y)$ is the scattering amplitude of an external parton that scatters on the wave function with color charge density $\rho$ (in our case - the projectile again). Thus the commutator in eq.(\ref{commut}) is the scattering amplitude of a dipole $(u,v)$ on the wave function created by the dipole $(x,y)$. In general, this scattering amplitude is an operator with nontrivial dependence on $\rho$ which provides for possibility of further scatterings. Thus for example an expression like 
 $P(zw)P(xy)\bar P(uv)$ corresponds to the double scattering of the $(u,v)$ dipole on the wave function created by two dipoles - $(x,y)$ and $(z,w)$. According to eq.(\ref{commut}) this expression would vanish. This simply corresponds to an approximation where any dipole can only scatter on one dipole at a time, or approximation of dilute projectile. If the projectile itself is dense, then of course multiple scatterings are allowed and the scattering amplitude of the external dipole $(u,v)$ is much more complicated. 
 
 Our aim now is to show why the approximation of eq. (\ref{aprbar}) is appropriate for the regime discussed in the present paper.
Consider again eq.(\ref{exp}) inserted in eq.(\ref{evoldip}) within the limit (\ref{max}). 
As we have explained, the leftmost exponential factor corresponds to the range of rapidities at which the object to the left of it (the projectile) is dilute, and the object to the right of it (evolved target) is dense, while the rightmost exponential factor acts in the opposite environment where the target is dilute, but the projectile is dense. In the leftmost factor therefore $H_{RFT}$ is approximated by 
$H_{KLWMIJ}$, while in the rightmost one by $H_{JIMWLK}$. Consider a generic term in expansion of the two exponentials to some order in $\Delta Y $:
\begin{equation}\label{expression}
\int d\alpha\int d\rho\delta(\rho)W_P[R]\Big(H_{KLWMIJ}[\rho,R]\Big)^n\Big(H_{JIMWLK}[\rho,R]\Big)^m\,e^{ig^2\int_x\rho^a(x)\alpha^a(x)}W_T[\alpha]
\end{equation}

In general we cannot expand $H_{KLWMIJ}$ in this expression in powers of $\delta/\delta\rho$, since any given term acts on a dense wave function to the right of it. By the same token $H_{JIMWLK}$ cannot be expanded in $\rho$, since it encounters a dense wave function when acting to the left. However, what interests us is how the terms involving $\bar P$ in $H_{JIMWLK}$ act on the $P$ operators in $H_{KLWMIJ}$. Technically, in order to calculate the integral in eq.(\ref{expression}), one has to commute all the factors involving $\rho$ all the way to the left, where they annihilate the $\delta$-function. Let us concentrate on a factor involving $S$, or equivalently $\bar P$ coming from one of the $H_{JIMLWK}$ terms. First, it has to be commuted through the rest of the factors $H_{JIMWLK}$ to the left of it. This is the rapidity interval where, according to our previous discussion, the projectile is dense and no further simplifications are possible. There is no problem however, since $H_{JIMWLK}$ is expressed in terms of $\bar P$ and $\bar P^\dagger$, and these operators have simple commutation relations, so the calculation is straightforward. The potential problem only arises when one has to commute $\bar P$ through the factors of $H_{KLWMIJ}$. However, once $\bar P$ arrives next to these factors, it is now in the environment where the projectile is dilute. Here $\bar P$ can be expanded to leading order in $\rho$. As a consequence, the commutation  relation of any factor $\bar P$ from $H_{JIMWLK}$ with any factor of $P$ from $H_{KLWMIJ}$ in eq. (\ref{expression}) can be approximated by the simple c-number relation eq. (\ref{commut}). 

The physics of this argument is simple. We can attribute the factor $\bar P$ that we have concentrated on, to the target wave function. Its interpretation then is that of the scattering amplitude of a dipole that belongs to the target, on the wave function of the evolved projectile. Since the projectile is dense, multiple scatterings are certainly possible, and therefore one cannot expand $\bar P$. However the structure of the evolved projectile wave function is such, that most of the gluons are soft. We know that from the exponential growth with rapidity of the gluon density in the dipole model . Thus the multiple scatterings will occur with large probability only on gluons close to the target rapidity (contained in $H_{JIMWLK}$ terms in eq.(\ref{expression}). On the other hand, there are not many gluons  at rapidities close to the "valence" rapidity of the projectile. Thus multiple scattering of the target dipole on these energetic gluons is suppressed by powers of the coupling constant. Therefore only single scattering term can be kept in the commutator of $\bar P$ close to target valence rapidity with $P$ close to the projectile valence rapidity. This is the approximation explored in ref. \cite{msw}.

\begin{figure}
\begin{center}
\includegraphics[scale=0.7]{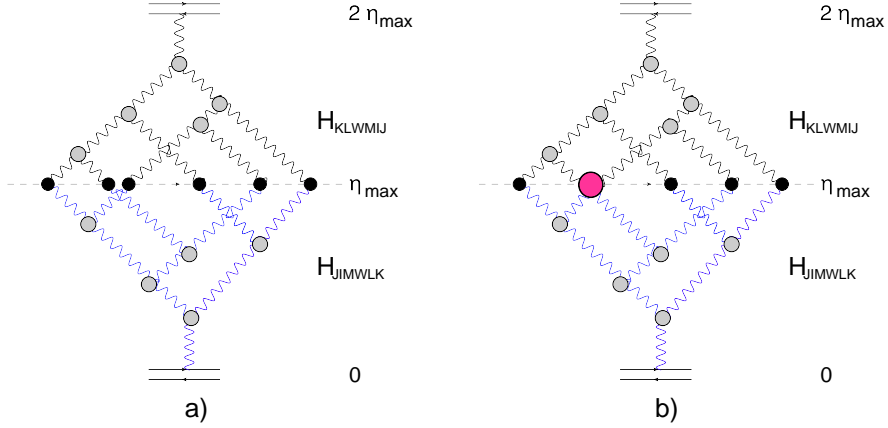}
\end{center}
\caption{\label{cascade} The JIMWLK and KLWMIJ Pomeron cascades described by $H_{RFT(P)}$ of \protect\eq{simple}. The wavy lines denote the BFKL Pomerons. The gray circles are the triple Pomeron vertex while the black circle denote  $\alpha_s^2\frac{1}{\nabla^2_1 \,\nabla^2_2}$ which is the amplitude of two dipoles interaction in the Born approximation of perturbative QCD. \fig{cascade}-b shows the first correction to $H_{RFT(P)}$ stemming from the induced four Pomeron vertex which describes the interaction of two dipoles from KLWMIJ cascade with two dipoles from JIMWLK cascade at low energy denoted by the large (red) circle.}
\end{figure}

The upshot of this discussion is that in the large Pomeron loop regime the RFT Hamiltonian (truncated to contain only Pomerons) can be written as (see \fig{cascade}-a)
\begin{equation}\label{simple}
H_{RFT(P)}=-\frac{\alpha_sN_c}{2\,\pi^2}\,\int_{x,y,z} M_{x,y;z}\,\Big\{[P_{x,z}+P_{z,y}\,-\,P_{x,y}]\,P^\dagger_{x,y}-P_{x,z}P_{z,y}\,P^\dagger_{x,y}-\bar P_{x,z}
\bar P_{z,y}\,\bar P^\dagger_{x,y}\Big\}
\end{equation}
with 
\begin{eqnarray}\label{bardagger}
\bar P(xy)&=&\frac{g^4}{4}[\frac{1}{\nabla^2}(xu)- \frac{1}{\nabla^2}(yu)][\frac{1}{\nabla^2}(xv)- \frac{1}{\nabla^2}(yv)][P^\dagger(uv)-\delta_{uv}\int_zP^\dagger_{uz}]\nonumber\\
\bar P^\dagger(xy)&=& \frac{4}{g^4}\nabla^2_x\nabla^2_yP(x,y), \ \ x\ne y; \ \ \ \ \ \bar P^\dagger(xy)=0, \ x=y
\end{eqnarray}
With the identification eq.(\ref{bardagger}) the Hamiltonian can be written in terms of the dual Pomerons in an explicitly self dual form (see Appendix A)
\begin{eqnarray}\label{simple1}
H_{RFT(P)}&=& -\frac{N_c}{32\,\pi^4\alpha_s}\int_{x,y,u,v} \nabla^2_u\nabla^2_v\,P_{u,v}
\left\{ \left[\frac{1}{\nabla^2}(xv)\,-\,\frac{1}{\nabla^2}(yv)\right]^2 \, 
\left[\frac{1}{ \nabla^2}(ux)+\frac{1}{ \nabla^2}(uy)\right]\right. +\nonumber \\
&&~~~~~~~~+\left. \left[\frac{1}{\nabla^2}(xu)\,-\,\frac{1}{\nabla^2}(yu)\right]^2 \, 
\left[\frac{1}{ \nabla^2}(vx)+\frac{1}{ \nabla^2}(vy)\right]
\right\} \,
\nabla^2_x\nabla^2_y\,\bar P_{x,y}\nonumber\\
&&+\frac{N_c}{8\,\pi^4\alpha_s}\,\int_{x,y,z} M_{x,y;z}\Big[P_{x,z}P_{z,y}\,\nabla^2_x\nabla^2_y\,\bar P_{x,y}+\bar P_{x,z}
\bar P_{z,y}\,\nabla^2_x\nabla^2_y\, P_{x,y}\Big]
\end{eqnarray}
This is  the Hamiltonian proposed by Braun in \cite{braun} and frequently referred to as BFKL Pomeron Calculus. 

\subsection{Beyond BFKL Pomeron Calculus}
The discussion in the previous subsection was not complete, since we have only considered the Pomeron contribution to the Hamiltonian. However, even at large $N_c$ one has to include other Reggeons, as explained in \cite{last}. The number of these Reggeons is in principle infinite, corresponding to all possible $n$-point functions of the matrix $R$. We will limit ourselves, like in \cite{last} to considering four lowest reggeons, and will include their contributions to $H_{RFT}$.
The Reggeons in questions are the Odderon, defined as 
\begin{equation}\label{odderon}
O(x,y)=\frac{1}{2N_c}\left({\rm tr} [R(x)R^\dagger(y)]-{\rm tr} [R(y)R^\dagger(x)]\right)\end{equation}
and the B and C reggeons defined in terms of the ``quadrupole'' operator
\begin{equation}\label{quadrupole}
Q(1,2,3,4)\equiv Q(x,y,u,v)\equiv \frac{1}{N_c}tr[R(x)R^\dagger(y)R(u)R^\dagger(v)]
\end{equation}
as
\begin{eqnarray}\label{BC-reggeon}
B(1,2,3,4)&=&\frac{1}{4}\left[4-Q(1,2,3,4)-Q(4,1,2,3)-Q(3,2,1,4)-Q(2,1,4,3)\right]\nonumber\\
&-&\left[P_{12}+P_{14}+P_{23}+P_{34}-P_{13}-P_{24}\right]\nonumber\\
C(1,2,3,4)&=&\frac{1}{4}\left[Q(1,2,3,4)+Q(4,1,2,3)-Q(3,2,1,4)-Q(2,1,4,3)\right]
\end{eqnarray}
As discussed in \cite{last}, the field $O$ is charge conjugation and signature odd, $C$ is signature even and charge conjugation odd, while $B$ has the same quantum numbers as the Pomeron.  When these fields are included, the KLWMIJ Hamiltonian has the form:
\begin{equation}\label{klwmijnew}
H_{KLWMIJ}=H_P+H_O+H_B+H_C
\end{equation}
where
\begin{eqnarray}\label{p}
H_P &=&-\frac{\alpha_sN_c}{2\,\pi^2}\,\int_{x,y,z} M_{x,y;z}\,\Big\{[P_{x,z}+P_{z,y}\,-\,P_{x,y}-P_{x,z}P_{z,y}-O_{x,z}O_{z,y}]\,P^\dagger_{x,y}\Big\}\\
H_O&=&-\frac{ \alpha_sN_c}{2\,\pi^2}\,\int_{x,y,z} M_{x,y;z}\,\Big\{[O_{x,z}+O_{z,y}\,-\,O_{x,y}-O_{x,z}P_{z,y}-P_{x,z}O_{z,y}]\,O^\dagger_{x,y}\Big\}
\end{eqnarray}
\begin{eqnarray}\label{hb}
H_C&=&-\frac{\alpha_sN_c}{2\,\pi^2}\,\int_{x,y,u,v,z}\Bigg\{-\,[M_{x,y;z}\,+\,M_{u,v;z}\,-\,L_{x,u,v,y;z}]
\,\,C_{xyuv}C^\dagger_{xyuv}+4L_{x,v,u,v;z}C_{xyuz}C^\dagger_{xyuv}\nonumber \\
&-&4L_{x,v,u,v;z}C_{xyuz}P_{zv}C^\dagger_{xyuv}\Bigg\}\\
&&\nonumber \\
H_B&=&-\frac{ \alpha_sN_c}{2\,\pi^2}\,\int_{xyuvz}\Bigg\{-\,[M_{x,y;z}\,+\,M_{u,v;z}\,-\,L_{x,u,v,y;z}]
\,\,B_{xyuv}B^\dagger_{xyuv}+4L_{x,v,u,v;z}B_{xyuz}B^\dagger_{xyuv}\nonumber\\
&-&2L_{x,y,u,v;z}\Big[P_{xv}P_{uy}+O_{xv}O_{uy}\Big]B^\dagger_{xyuv}\nonumber\\
&-&2P_{xz}P_{yz}\Big[2L_{x,y,u,v;z}B^\dagger_{xyuv}-\Big(L_{x,u,y,v;z}+L_{x,v,y,u;z}\Big)B^\dagger_{xuyv}\Big]\nonumber\\
&-&4P_{xz}P_{yu}\Big[2L_{x,y,x,v;z}B^\dagger_{xyuv}-L_{x,y,x,u;z}B^\dagger_{xyvu}\Big]
-4B_{xyuz}P_{zv}L_{x,v,u,v;z}B^\dagger_{xyuv}\Bigg\}
\end{eqnarray}
where
\begin{eqnarray}
L_{x,y,u,v;z}\,&=&\,\left[\frac{(x\,-\,z)_i}{ (x\,-\,z)^2}\,-\,\frac{(y\,-\,z)_i}{(y\,-\,z)^2}\right ]\,\,
\left[\frac{(u\,-\,z)_i}{(u\,-\,z)^2}\,-\,\frac{(v\,-\,z)_i}{(v\,-\,z)^2}\right ]\nonumber \\
&=&\,\,\frac{1}{ 2}\,\left[\,M_{y,u;z}\,+\,M_{x,v;z}\,-\,M_{y,v;z}-M_{x,u;z}\,\right]
\end{eqnarray}
Applying the DDD transformation, we can write immediately the JIMWLK Hamiltonian in terms of the dual operators $\bar O$, $\bar B$ and $\bar C$ obtained by 
replacing $R$ by $S$ in all expressions eqs. (\ref{odderon},\ref{quadrupole}, \ref{BC-reggeon}).
\begin{equation}\label{klwmijnew1}
H_{JIMWLK}=H_{\bar P}+H_{\bar O}+H_{\bar B}+H_{\bar C}
\end{equation}
where
\begin{eqnarray}\label{p1}
H_{\bar P} &=&-\frac{\alpha_sN_c}{2\,\pi^2}\,\int_{x,y,z} M_{x,y;z}\,\Big\{[\bar P_{x,z}+\bar P_{z,y}\,-\,\bar P_{x,y}-\bar P_{x,z}\bar P_{z,y}-\bar O_{x,z}\bar O_{z,y}]\,\bar P^\dagger_{x,y}\Big\}\\
H_{\bar O}&=&-\frac{\alpha_sN_c}{2\,\pi^2}\,\int_{x,y,z} M_{x,y;z}\,\Big\{[\bar O_{x,z}+\bar O_{z,y}\,-\,\bar O_{x,y}-\bar O_{x,z}\bar P_{z,y}-\bar P_{x,z}\bar O_{z,y}]\,\bar O^\dagger_{x,y}\Big\}
\end{eqnarray}
\begin{eqnarray}\label{hbarb}
H_{\bar C}&=&-\frac{\alpha_s N_c}{2\,\pi^2}\int_{x,y,u,v,z}\Bigg\{-[M_{x,y;z}+M_{u,v;z}-L_{x,u,v,y;z}]
\bar C_{xyuv}C^\dagger_{xyuv}+4L_{x,v,u,v;z}\bar C_{xyuz}\bar C^\dagger_{xyuv}\nonumber \\
&-&4L_{x,v,u,v;z}\bar C_{xyuz}\bar P_{zv}\bar C^\dagger_{xyuv}\Bigg\}\\
&&\nonumber \\
H_{\bar B}&=&-\frac{\alpha_sN_c}{2\,\pi^2}\,\int_{xyuvz}\Bigg\{-\,[M_{x,y;z}\,+\,M_{u,v;z}\,-\,L_{x,u,v,y;z}]
\,\,\bar B_{xyuv}\bar B^\dagger_{xyuv}+4L_{x,v,u,v;z}\bar B_{xyuz}\bar B^\dagger_{xyuv}\nonumber\\
&-&2L_{x,y,u,v;z}\Big[\bar P_{xv}\bar P_{uy}+\bar O_{xv}\bar O_{uy}\Big]\bar B^\dagger_{xyuv}\nonumber\\
&-&2\bar P_{xz}\bar P_{yz}\Big[2L_{x,y,u,v;z}\bar B^\dagger_{xyuv}-\Big(L_{x,u,y,v;z}+L_{x,v,y,u;z}\Big)\bar B^\dagger_{xuyv}\Big]\nonumber\\
&-&4\bar P_{xz}\bar P_{yu}\Big[2L_{x,y,x,v;z}\bar B^\dagger_{xyuv}-L_{x,y,x,u;z}\bar B^\dagger_{xyvu}\Big]
-4\bar B_{xyuz}\bar P_{zv}L_{x,v,u,v;z}\bar B^\dagger_{xyuv}\Bigg\}
\end{eqnarray}

The same line of argument as in the previous subsection leads us to combine these two sets Hamiltonians into $H_{RFT}$ with an additional simplification that the commutation relations between 
$P,\ O,\ C, \ B$ and $\bar P,
 \bar O, \bar C, \bar B$ are taken to be those of the dilute limit. To establish these commutation relations we have to expand the dual reggeons to leading order in 
 $\rho$ and relate them to the conjugate reggeons.

For the expansion of $B$ and $\bar B$ reggeons we have
\begin{eqnarray}\label{bnc}
B(x,y,u,v)&=&\frac{1}{4N_c}{\rm Tr}\{[T^aT^bT^cT^d]+[T^dT^cT^bT^a]\}\times\\
&\times&(\frac{\delta}{\delta \rho^a(x)}-\frac{\delta}{\delta \rho^a(v)})(\frac{\delta}{\delta \rho^b(x)}-\frac{\delta}{\delta \rho^b(y)})
(\frac{\delta}{\delta \rho^c(y)}-\frac{\delta}{\delta \rho^c(u)})(\frac{\delta}{\delta \rho^d(u)}-\frac{\delta}{\delta \rho^d(v)})\nonumber
\end{eqnarray}
\begin{eqnarray}\label{bbarnc}
\bar B(\bar x,\bar y,\bar u,\bar v)&=&\frac{g^8}{4N_c}{\rm Tr}\{[T^aT^bT^cT^d]+[T^dT^cT^bT^a]\}\times\\
&\times&(\alpha^a(\bar x)-\alpha^a(\bar v))(\alpha^b(\bar x)-\alpha^b(\bar y))
(\alpha^c(\bar y)- \alpha^c(\bar u))(\alpha^d(\bar u)-\alpha^d(\bar v))\nonumber
\end{eqnarray}
The commutator is
\begin{eqnarray}\label{commutb}
[\bar B(\bar x,\bar y,\bar u,\bar v),B(x,y,u,v)]&=&\frac{g^8N^2_c}{16\cdot 8}\,{\cal D}_B(\bar x,\bar y,\bar u,\bar v;x,y,u,v)\,; \nonumber \\
{\cal D}_B(\bar x,\bar y,\bar u,\bar v;x,y,u,v)&\equiv&\Bigg[\Delta(\bar x, \bar v; x,v)
\Delta(\bar x, \bar y; x,y)\Delta(\bar y, \bar u; y,u)\Delta(\bar u, \bar v; u, v)\nonumber\\
&+&\Delta(\bar x, \bar v; x,v)
\Delta(\bar x, \bar y; u,v)\Delta(\bar y, \bar u; y,u)\Delta(\bar u, \bar v; x,y)\nonumber\\
&+&\Delta(\bar x, \bar v; x,y)
\Delta(\bar x, \bar y; y,u)\Delta(\bar y, \bar u; u, v)\Delta(\bar u, \bar v; x,v)\nonumber\\
&+&\Delta(\bar x, \bar v; x,y)
\Delta(\bar x, \bar y; x,v)\Delta(\bar y, \bar u; u, v)\Delta(\bar u, \bar v;y,u )\nonumber\\
&+&\Delta(\bar x, \bar v; y,u)
\Delta(\bar x, \bar y; u, v)\Delta(\bar y, \bar u;x,v )\Delta(\bar u, \bar v; x,y)\nonumber\\
&+&\Delta(\bar x, \bar v; y,u)
\Delta(\bar x, \bar y; x,y)\Delta(\bar y, \bar u;x,v )\Delta(\bar u, \bar v; u, v)\nonumber\\
&+&\Delta(\bar x, \bar v;u, v )
\Delta(\bar x, \bar y; x,v)\Delta(\bar y, \bar u;x,y )\Delta(\bar u, \bar v;y,u )\nonumber\\
&+&\Delta(\bar x, \bar v;u, v )
\Delta(\bar x, \bar y; y,u)\Delta(\bar y, \bar u;x,y )\Delta(\bar u, \bar v;x,v )\Bigg]
\end{eqnarray}

An entirely analogous calculation can be performed for the $C$-reggeon, with the only difference that $[T^aT^bT^cT^d]+[T^dT^cT^bT^a]\rightarrow [T^aT^bT^cT^d]-[T^dT^cT^bT^a]$ in eqs. (\ref{bnc}) and (\ref{bbarnc}). The result for the commutator is identical with eq.(\ref{commutb})
\begin{eqnarray}\label{commutc}
[\bar C(\bar x,\bar y,\bar u,\bar v),C(x,y,u,v)]&=&\frac{g^8N^2_c}{16\cdot 8}\Bigg[\Delta(\bar x, \bar v; x,v)
\Delta(\bar x, \bar y; x,y)\Delta(\bar y, \bar u; y,u)\Delta(\bar u, \bar v; u, v)\nonumber\\
&+&\Delta(\bar x, \bar v; x,v)
\Delta(\bar x, \bar y; u,v)\Delta(\bar y, \bar u; y,u)\Delta(\bar u, \bar v; x,y)\nonumber\\
&+&\Delta(\bar x, \bar v; x,y)
\Delta(\bar x, \bar y; y,u)\Delta(\bar y, \bar u; u, v)\Delta(\bar u, \bar v; x,v)\nonumber\\
&+&\Delta(\bar x, \bar v; x,y)
\Delta(\bar x, \bar y; x,v)\Delta(\bar y, \bar u; u, v)\Delta(\bar u, \bar v;y,u )\nonumber\\
&+&\Delta(\bar x, \bar v; y,u)
\Delta(\bar x, \bar y; u, v)\Delta(\bar y, \bar u;x,v )\Delta(\bar u, \bar v; x,y)\nonumber\\
&+&\Delta(\bar x, \bar v; y,u)
\Delta(\bar x, \bar y; x,y)\Delta(\bar y, \bar u;x,v )\Delta(\bar u, \bar v; u, v)\nonumber\\
&+&\Delta(\bar x, \bar v;u, v )
\Delta(\bar x, \bar y; x,v)\Delta(\bar y, \bar u;x,y )\Delta(\bar u, \bar v;y,u )\nonumber\\
&+&\Delta(\bar x, \bar v;u, v )
\Delta(\bar x, \bar y; y,u)\Delta(\bar y, \bar u;x,y )\Delta(\bar u, \bar v;x,v )\Bigg]
\end{eqnarray}
A similar calculation gives for the Odderon
\begin{eqnarray}\label{ocomm}
[O(xy),\bar O(uv)]=\frac{g^6N_c}{32}\Big[&&2\frac{1}{\nabla^2}(uy)\frac{1}{\nabla^2}(vy)\frac{1}{\nabla^2}(vx)+2\frac{1}{\nabla^2}(uy)\frac{1}{\nabla^2}(ux)\frac{1}{\nabla^2}(vx)\nonumber\\
&&-2\frac{1}{\nabla^2}(uy)\frac{1}{\nabla^2}(vy)\frac{1}{\nabla^2}(ux)-2\frac{1}{\nabla^2}(ux)\frac{1}{\nabla^2}(vy)\frac{1}{\nabla^2}(vx)\nonumber\\
&&+\frac{1}{\nabla^2}(ux)\frac{1}{\nabla^2}(vy)\frac{1}{\nabla^2}(vy)
+\frac{1}{\nabla^2}(vy)\frac{1}{\nabla^2}(ux)\frac{1}{\nabla^2}(ux)\nonumber\\
&&-\frac{1}{\nabla^2}(vx)\frac{1}{\nabla^2}(uy)\frac{1}{\nabla^2}(uy)-\frac{1}{\nabla^2}(uy)\frac{1}{\nabla^2}(vx)\frac{1}{\nabla^2}(vx)\Big]
\end{eqnarray}

It is clear therefore that the extension of the pure Pomeron Hamiltonian, which also accounts for higher Reggeons in the "Pomeron loop" regime
should be taken as the sum 
\begin{equation}\label{rft}
H_{RFT}=H_{RFT(P)}+H_{B+\bar B}+H_{C+\bar C}+H_{O+\bar O}
\end{equation}
with
\begin{eqnarray}\label{b+barb}
H_{B+\bar B}&=&-\frac{\alpha_sN_c}{2\pi^2}\int_{xyuvz}\Big\{-[M_{x,y;z}+M_{u,v;z}-L_{x,u,v,y;z}]
B_{xyuv}B^\dagger_{xyuv}+4L_{x,v,u,v;z}B_{xyuz}B^\dagger_{xyuv}\nonumber\\
&-&2L_{x,y,u,v;z}\Big[P_{xv}P_{uy}+O_{xv}O_{uy}\Big]B^\dagger_{xyuv}-2L_{x,y,u,v;z}\Big[\bar P_{xv}\bar P_{uy}+\bar O_{xv}\bar O_{uy}\Big]
\bar B^\dagger_{xyuv}\nonumber\\
&-&2P_{xz}P_{yz}\Big[2L_{x,y,u,v;z}B^\dagger_{xyuv}-\Big(L_{x,u,y,v;z}+L_{x,v,y,u;z}\Big)B^\dagger_{xuyv}\Big]\nonumber\\
&-&2\bar P_{xz}\bar P_{yz}\Big[2L_{x,y,u,v;z}\bar B^\dagger_{xyuv}-\Big(L_{x,u,y,v;z}+L_{x,v,y,u;z}\Big)\bar B^\dagger_{xuyv}\Big]\nonumber\\
&-&4P_{xz}P_{yu}\Big[2L_{x,y,x,v;z}B^\dagger_{xyuv}-L_{x,y,x,u;z}B^\dagger_{xyvu}\Big]
-4B_{xyuz}P_{zv}L_{x,v,u,v;z}B^\dagger_{xyuv}\nonumber\\
&-&4\bar P_{xz}\bar P_{yu}\Big[2L_{x,y,x,v;z}\bar B^\dagger_{xyuv}-L_{x,y,x,u;z}\bar B^\dagger_{xyvu}\Big]
-4\bar B_{xyuz}\bar P_{zv}L_{x,v,u,v;z}\bar B^\dagger_{xyuv}\Big\}
\end{eqnarray}
\begin{eqnarray}\label{c+barc}
H_{C+\bar C}&=&-\frac{ \alpha_sN_c}{2\pi^2}\int_{x,y,u,v,z}\Big\{-[M_{x,y;z}+M_{u,v;z}-L_{x,u,v,y;z}]
C_{xyuv}C^\dagger_{xyuv}+\\
&+&4L_{x,v,u,v;z}C_{xyuz}C^\dagger_{xyuv}-4L_{x,v,u,v;z}C_{xyuz}P_{zv}C^\dagger_{xyuv}
-4L_{x,v,u,v;z}\bar C_{xyuz}\bar P_{zv}\bar C^\dagger_{xyuv}
\Big\}\nonumber
\end{eqnarray}
\begin{eqnarray}\label{o_baro}
H_{O+\bar O}&=&-\frac{\alpha_sN_c}{2\,\pi^2}\,\int_{x,y,z} M_{x,y;z}\,\Bigg[
\Big\{[O_{x,z}+O_{z,y}\,-\,O_{x,y}\Big\}\,O^\dagger_{x,y}\nonumber\\
&&-[O_{x,z}P_{z,y}+P_{x,z}O_{z,y}]\,O^\dagger_{x,y}
-[\bar O_{x,z}\bar P_{z,y}+\bar P_{x,z}\bar O_{z,y}]\,\bar O^\dagger_{x,y}\Bigg]
\end{eqnarray}
This is supplemented with the commutation relations eqs.(\ref{commutb},\ref{commutc},\ref{ocomm}).

Just like in the case of the Pomeron, one can get rid of $B^\dagger$ and $C^\dagger$ in favor of the dual reggeons $\bar B$ and $\bar C$ in the Hamiltonian, and rewrite the relevant terms in an explicitly self dual form. 
To do this, we recall that the conjugate B-reggeon to leading order is given by the expression (\cite{last})
\begin{equation}
B^\dagger(1234)=-\frac{2}{N_c^3}{\rm tr}(T^aT^bT^cT^d)\Big\{\rho^a(1)\rho^b(2)\rho^c(3)\rho^d(4)+\rho^a(2)\rho^b(1)\rho^c(4)\rho^d(3)\Big\}
\end{equation}
Using this, and a similar relation for $C^\dagger$, to leading order we have 
\begin{eqnarray}\label{bcbardagger}
\nabla^2_x\nabla^2_y\nabla^2_u\nabla^2_v\bar B(x,y,u,v)&=&\frac{g^8N_c^2}{4}B^\dagger(x,y,u,v);\nonumber \\
\nabla^2_x\nabla^2_y\nabla^2_u\nabla^2_v\bar C(x,y,u,v)&=&\frac{g^8N_c^2}{4}C^\dagger(x,y,u,v)
\end{eqnarray}

The situation is more subtle for the Odderon, since we do not have readily available an equation similar to eq.(\ref{bcbardagger}). We therefore cannot explicitly eliminate $O^\dagger$ in favor of $\bar O$. However from practical point of view this is not necessary, since the commutation relation eq.(\ref{ocomm}) provides all the necessary information to be able to use $O$ and $\bar O$ within the same framework.

We conclude this section with discussion of two important aspects of the Hamiltonian $H_{RFT}$.

\subsection {On the peculiarities of large $N_c$ counting.}

There is a certain subtlety related to the Hamiltonian eq.(\ref{b+barb}) that we have to comment upon. As we have discussed above, to write down $H_{RFT}$ we had to add the JIMWLK and KLWMIJ Hamiltonians, and subtract the BFKL term. When performing this in terms of the B-reggeon operators, to arrive at eq.(\ref{b+barb}) we have added $H_B$ and $H_{\bar B}$ and subtracted the homogeneous term, which is common to the two Hamiltonians, while keeping all vertices intact. On the other hand, as noted in \cite{last}, the $B^\dagger PP$ vertex in eq.(\ref{hb}) in the expansion to lowest order in $\delta/\delta\rho$ generates a contribution to the linear BKP equation, and thus in a sense is a part of the BFKL Hamiltonian. The same goes for the vertex $\bar B^\dagger \bar P\bar P$ in eq.(\ref{hbarb}). This begs the question, whether we have not undersubtracted the BFKL terms in arriving to eq.(\ref{b+barb}) by keeping both these vertices. It does of course seem very unnatural to subtract either or both of these vertices, and in fact they are clearly both needed so that $H_{B+\bar B}$ can reproduce all the terms in $H_{KLWMIJ}$ in the dilute regime, and all the terms in $H_{JIMWLK}$ in the dense one. 

It is necessary and consistent to keep both these terms in the Hamiltonian, if and only if we can show that in the regime they are not supposed to be present, they are parametrically smaller than the terms we have subtracted. We will now show that this is indeed the case due to the somewhat peculiar way the large $N_c$ limit works at high energy. Specifically, we will show that the term  $B^\dagger PP$ is leading order at large $N_c$ in the KLWMIJ regime and thus is the same order as the homogeneous term $B^\dagger B$. On the other hand in the dense JIMWLK regime the $B^\dagger PP$ term is suppressed in the large $N_c$ limit, even though naively one could think that it is always $O(1)$. Thus even though we have ``undersubtracted'' this term in the JIMWLK regime, this is consistent within our calculation, since we are working in the large $N_c$ limit.  The converse is true for the $\bar B^\dagger \bar P\bar P$ term. It is $O(1)$ in the JIMWLK regime, but $O(1/N_c^2)$ in the KLWMIJ regime.
Thus keeping both vertices in the Hamiltonian is a completely consistent approximation in the large $N_c$ limit.

To understand the peculiarities of the large $N_c$ counting, consider for example the dual Pomeron amplitude eq. (\ref{dualP}).
The saturation regime is defined as regime where the matrix $S$ has fluctuation of order unity, and therefore close to the saturation regime in terms of $1/N_c$ counting, 
$\bar P(x,y)\sim O(1)$. 

Now let us consider the dilute regime. To count powers of $N_c$ properly, we have to restore the powers of the coupling constant. Recall that 
$S=\exp\{ig^2T^a\alpha^a(x)\}$, where $\alpha$ is the gauge field in the wave function of the projectile. The gauge field itself is determined by the color charge density  via the Yang-Mills equation as $\nabla^2\alpha=\rho$, and so $S(x)=\exp\{ig^2T^a\nabla^{-2}\rho^a\}$. In the last relation the normalization of the color charge density is such that the charge for a single particle is of order unity, or more precisely the second Casimir operator in a fundamental representation is equal to $N_c/2$ without any powers of the coupling constant $g$. Thus in the dilute  regime, where we assume that the projectile contains a finite number of partons, we find
\begin{equation}
\bar P(x,y)\sim \frac{g^4}{4N_c}[\nabla^{-2}(x,u)-\nabla^{-2}(y,u)][\nabla^{-2}(x,v)-\nabla^{-2}(y,v)]\rho^a(u)\rho^a(v)\propto \alpha_s^2
\end{equation}
In the usual 't Hooft counting, $\lambda\equiv\alpha_s\,N_c$  is finite in the large $N_c$ limit, $\alpha_s\propto \frac{1}{N_c}$  and thus 
\begin{equation}
\bar P\sim \frac{1}{N^2_c}
\end{equation}
This is natural, given the fact that in the large $N_c$ limit mesons (and therefore heavy quarkonia - dipoles) in QCD are stable, noninteracting particles.

Similarly, considering the Pomeron in the situation where the projectile scatters on a target that contain a finite number of partons (or dipoles), we find
\begin{equation}
P\sim \alpha_s^2\sim \frac{1}{N_c^2}
\end{equation}
This is obtained taking $\frac{\delta}{\delta\rho}\rightarrow g^2\nabla^2\rho$.
Interestingly, in this regime the B-Reggeon is parametrically
larger than the product of two Pomerons. Expanding the quadrupole operator
to order $(\frac{\delta}{\delta\rho})^4$ we find eq.(\ref{bnc}). The magnitude of this expression in the large $N_c$ limit is estimated taking again $\frac{\delta}{\delta\rho}\rightarrow g^2\nabla^2\rho$
and calculating the product of four generators of
SU(Nc) group in some low dimensional representation. For example, averaging over a fundamental representation we have
\begin{equation}
\langle\rho^a\rho^b\rho^c\rho^d\rangle_{fundamental}=\frac{1}{N_c}{\rm Tr} [T^aT^bT^cT^d]
\end{equation}
using this in eq.(\ref{bnc}) we find 
\begin{equation}
B(x,y,u,v)\sim \alpha_s^4N^2_c=\alpha^2_s\lambda^2=\lambda^4/N_c^2
\end{equation}
Note that in this regime the $B$ reggeon is of order $1/N_c^2$, while the two Pomeron contribution to scattering is $O(1/N^4_c)$; and thus $B\gg P^2$. 

This $1/N_c$ counting brings forth an interesting point. In order to get to the saturation regime at large $N_c$ one needs a parametrically large energy. The Pomeron amplitude becomes of order one only at rapidity determined by
\begin{equation}
\alpha_s^2\,e^{\,\omega_P Y}\sim 1; \ \ \ \ \ \ Y\sim \frac{1}{\omega_P} [\ln \frac{1}{\omega_P}+\ln N_c]
\end{equation}
Therefore the $N_c$ counting for the same scattering amplitude is very different in dense and dilute limits.

Returning to the estimates of various terms in the Hamiltonian $H_{B+\bar B}$ , for definiteness let us concentrate on the term $B^\dagger PP$. This term is certainly important in the KLWMIJ regime, where as we know, it is of the same order as the homogeneous term $B^\dagger B$, which is undoubtedly a part of $H_{BFKL}$. In this regime $PP\sim O(1)$ and also $B\sim O(1)$. On the other hand in the JIMWLK regime things are very different as is obvious from the previous discussion. The B-reggeon operator is of order $B\sim \alpha_s^4 N_c^2$, while the two Pomeron operator is parametrically smaller $PP\sim \alpha_s^4$. Thus in the JIMWLK regime the vertex we have "undersubtracted" is suppressed by $1/N_c^2$ relatively to the homogeneous term $B^\dagger B$. Therefore keeping it is perfectly consistent in the large $N_c$ limit, which is the approximation we are using in this paper. The reverse is true for the vertex $\bar B^\dagger \bar P\bar P$. It is as large as $BB^\dagger$ in the JIMWLK regime, but is suppressed by the factor $1/N_c^2$ in the KLWMIJ regime, and again, it is fully consistent to keep it in the Hamiltonian ``as is''. 
 
\subsection{On the effective $2\rightarrow 2$ Pomeron vertex }

A possibility that RFT should involve a $2\rightarrow 2$ Pomeron vertex has been previously discussed in the literature \cite{BALERY, levinlublinsky, brunet}. 
Recently, Braun \cite{braun..} argued that such a vertex appears in the BKP formalism  and can be relevant for collisions of two deuterons.

The Hamiltonian eq.(\ref{rft}) indeed gives rise to such an effective vertex
 via integration of an intermediate exchange of a Pomeron and of a B-reggeon. The latter exchange gives dominant contribution in the 't Hooft large $N_c$ limit.
 
The Hamiltonian eq.(\ref{rft}) contains the B-reggeon as an independent degree of freedom. In general this is necessary, since some physical observables involve $B$ directly, as for example double inclusive gluon production as discussed in the next section. However if we are interested in restricted set of observables which depend only on the Pomeron field, we can ``integrate out'' the $B$-reggeon and obtain the effective Pomeron Hamiltonian\footnote{We should also integrate of course the other reggeons, $O$ and $C$. These however give subleading contributions and we do not discuss them here.}.
This integration out procedure clearly generates an effective $2\rightarrow 2$ Pomeron vertex due to contributions of a single $B$-reggeon intermediate state. A similar contribution arises also from a single Pomeron intermediate state.

 Fig. 2 illustrates the effective four Pomeron vertex due to exchanges of the Pomeron and the B-Reggeon. 
 The fundamental vertices of $H_{RFT}$ that couple the two Pomerons to one $B$-Reggeon state  are $PPB^\dagger$ and $\bar P\bar P\bar B^\dagger$ in \eq{b+barb}, while coupling to one Pomeron state is via the three Pomeron vertex in \eq{simple}. 
 
\begin{figure}
\begin{center}
\includegraphics[scale=0.8]{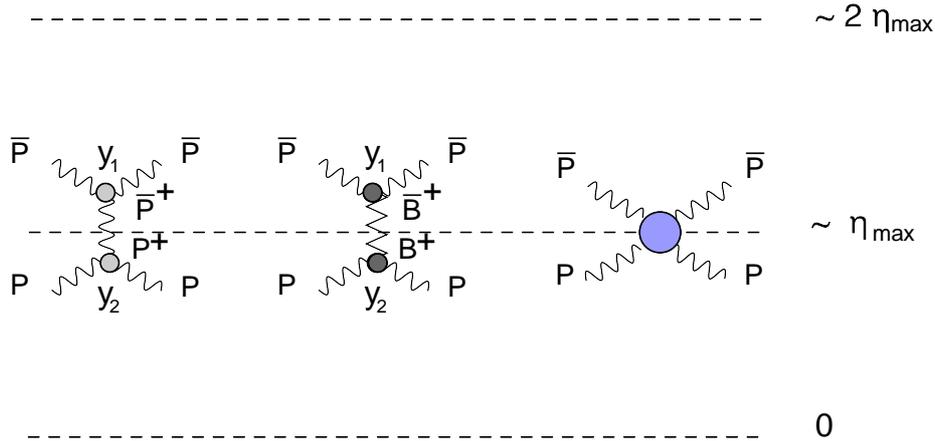}
\end{center}
 \caption{\label{4pom} The four Pomeron interactions that are  generated by the exchange of the Pomeron and B-Reggeon.}
  \end{figure}

 The physical meaning of the four Pomeron interaction as illustrated in \fig{4pom}, is that of the probability for ``direct'' interaction of two dipoles from the upper cascade with two dipoles from the lower cascade. For dilute-dense scattering this probability is small. However when the rapidity of one of the colliding objects approaches $2 \eta_{max}$ the four Pomeron interaction becomes more significant. It is the largest if all the dipoles that interact are close to midrapidity $\eta=\eta_{max}$.
 
  The parametric estimate for the strength of the direct four Pomeron interaction due to one Pomeron exchange is $\lambda^3/N^2_c$, while for the $B$-Reggeon it is $ \lambda/N^2_c$
The Pomeron exchange can therefore be neglected, or rather treated as a perturbation. 

To derive the induced interaction due to the $B$-Reggeon we need to integrate over the rapidity of the intermediate state. This gives a factor $1/\Lb 2\,\omega_{\mbox{\tiny P}}  - \omega_{\mbox{\tiny B}}\Rb$ where $\omega_{\mbox{\tiny P}}$ is the intercept of the BFKL Pomeron  and $\omega_{\mbox{\tiny B}} $ is the intercept of the B-Reggeon exchange. 
Thus the effect of $B$-reggeon propagation on Pomeron observables in the range of validity of $H_{RFT}$ can be summarized by the effective Hamiltonian
\bea \label{genh}
&&H_{ RFT(E)}\,\,=\,\,H_{RFT(P)}
+\,\,\frac{1}{\alpha_s^2}\frac{1}{8\pi^8}\frac{1}{ \Lb 2\,\omega_{\mbox{\tiny P}} - \omega_{\mbox{\tiny B}}\Rb }  \int_{x,y, u,v;  \bar{x},\bar{y}, \bar{u}, \bar{v}; z, \bar{z}}
\nonumber\\&&\,\,
\Bigg\{\Big( 4 P_{x z} \Lb P_{y z} - P_{u z}\Rb \,+\,2 P_{x v} P_{u y}\Big)\,L_{x,y,u,v;z} \,+\,8 P_{x z}\Lb P_{y v} \,-\,P_{y u}\Rb\,L_{x,y,x,v;z}\Bigg\} \nonumber\\
&& \times\,\,\,{\cal D}_B \Lb x, y, u, v;  \bar{x}, \bar{y}, \bar{u}, \bar{v}\Rb\,\nonumber\\
&& \times\,\,\Bigg\{\Big( 4 {\bar P}_{\bar x \bar z} \Lb \bar P_{\bar y \bar z} - \bar P_{\bar u \bar z}\Rb \,+\,2  \bar P_{ \bar x  \bar v}  \bar P_{ \bar u  \bar y}\Big)\,L_{ \bar x, \bar y, \bar u, \bar v; \bar z} \,+\,8  \bar P_{ \bar x \bar z}\Lb  \bar P_{ \bar y  \bar v} \,-\, \bar P_{ \bar y \bar u}\Rb\,L_{\bar x,\bar y,\bar x,\bar v; \bar z}\Bigg\}
\eea
where ${\cal D}_B$  
is defined in \eq{commutb}.

An interesting property of this Hamiltonian is that it generates the additional contribution to the two Pomeron Green's function\cite{BALERY} through graphs illustrated on Fig.3. This has the effect that in the linearized approximation the two Pomeron Green's function increases faster than the product of two the single Pomeron  exchanges: $\omega_{PP} \,\,=\,\,2\, \omega_{\mbox{\tiny P}} \,+\,\lambda/N^2_c \,\delta$ where $\delta$ is a small number (see Ref.\cite{BALERY} for details). This contribution starts to be essential with rapidities $\eta \propto N^2_c/\lambda \,>\,\eta_{max}$.

\begin{figure}
\begin{minipage}{8.5cm}{
\centerline{\includegraphics[scale=0.25]{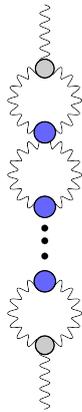}}}
\end{minipage}
\begin{minipage}{ 6.5cm}{
 \caption{\label{2pom} The contribution of  four Pomeron interactions to Green's function of two Pomerons. The coupling to a single Pomeron state is indicated by the upper and lower three Pomeron vertices.}}
 \end{minipage}
  \end{figure}

A word of caution is in order here.  While we have demonstrated the emergence of an effective $2\rightarrow 2$ Pomeron vertex from $H_{RFT}$, 
 the exact form of the vertex itself is not under control within our approximation. First, we have systematically ignored all the subleading $N_c$
 effects while deriving the RFT in this paper. On the other hand the two dipole-two dipole interaction, which is described by the effective vertex eq.(\ref{genh}) is subleading at large $N_c$. Thus we cannot exclude other sources contributing to this vertex beyond the ones taken into account in eq.(\ref{genh}). Second, as has been emphasized above, although we have written $H_{RFT}$ in the operator form, strictly speaking in its region of validity all the vertices close in rapidity to a dilute object must be of the splitting type. The   $2\rightarrow 2$ vertex is generated by the diagrams (Fig.2) which involve the ``wrong'' order of vertices: a merging vertex is closest to a dilute object, and a splitting appears only further away in rapidity. Such a diagram is subleading in the regime of validity of $H_{RFT}$ eq.(\ref{rft}). In fact such order of vertices naturally corresponds to collision of two dense objects. Indeed Braun argues \cite{braun..} 
 that the $2\rightarrow 2$ is enhanced in the dense regime. This is however the regime where, as we have discussed above our Hamiltonian is not under control. 
 Thus, although our considerations here regarding the existence of $2\rightarrow 2$ Pomeron vertex are plausible, it remains to be understood if they can be put on a firmer basis.

\section{Gluon production.}

The previous discussion pertained to the energy evolution of scattering amplitudes. Another set of interesting observables at high energy are inclusive (multi)gluon production amplitudes.
In the context of $H_{RFT}$ of ref.\cite{aklp} they were discussed in \cite{rftproduction}. The derivation of \cite{rftproduction} employed the same approximations as in 
\cite{aklp}, and thus has the same range of validity. In this section we discuss the same type of simplification for gluon production observables as the one discussed in Section 3 for $H_{RFT}$. We limit ourselves to discussing inclusive single gluon production and inclusive production of two gluons that have similar rapidity, so that the evolution between the rapidities of the gluons can be neglected. The rapidity at which the gluons are measured can be either close to the projectile, or close to the target, or anywhere in between. The restriction that follows from the derivation of \cite{rftproduction} is that at no rapidity is a dense multigluon final state produced in the scattering process.

As in previous sections, our goal is to cast the gluon production observables in terms of the natural degrees of freedom of Reggeon Field Theory, which are the dual pairs $P$ and $\bar P$; $B$ and $\bar B$ and so on. 
We start with discussing the simplest observable of this type - a single gluon inclusive production.

\subsection{Single Gluon Production.}
Consider inclusive production of a gluon with rapidity $\eta$ in a scattering process at total energy corresponding to rapidity difference $Y$ between the target and the projectile. First, we take $\eta$ to be close enough to the projectile, so that the projectile wave function evolved to $\eta$ is dilute. We do not assume diluteness of the target wave function. In this regime the amplitude of \cite{rftproduction} reduces to a well know expression first derived in \cite{kovtuch}. We find it more convenient to use the notations of \cite{gluons,KLW}. The single gluon production amplitude according to \cite{gluons,KLW} is given by 
\begin{eqnarray}\label{singleg}
\frac{d\sigma}{ d\eta\,dk^2}&=&\frac{\alpha_s}{4\,\pi^3}\,
\int_{b;z,\bar z} e^{i\,k(z\,-\,\bar z)}\,\int_{x,y} \frac{(z-x)_i}{(z-x)^2}\,
\frac{(\bar z-y)_i}{(\bar z-y)^2} \,\times\nonumber \\
&\times&\langle W_T^{ Y-\eta}\vert
\left[(S_z^{A\dagger}\,-\,S_x^{A\dagger})( S^A_{\bar z}\,-\, S^A_y)\right]^{ab}\rangle\langle W_P^\eta\vert J_L^a[x]\,J_R^b[y]\rangle
\end{eqnarray}
The target matrix elements are defined as
\begin{equation}
\langle W_T^{ Y-\eta}\vert O[S]\rangle\equiv \int D\rho_T\, W_T^{\bar Y-\eta}[\rho_T] \ O[S]; 
\end{equation} 
and similarly for the projectile. In these expressions the target distribution $W^T$ has been evolved through the rapidity interval of length $Y-\eta$ , while the projectile distribution $W_P$ has been evolved by $\eta$ to the rapidity of the observed gluon.

We have explicitly indicated in eq.(\ref{singleg}) the integration over the impact parameter $\vec b$, which is the transverse plane vector  between the ``center of mass'' of the projectile and the target wave functions. One of the wave functions ($W_T$ or $W_P$) should be understood as depending on $\vec b$ through a global shift of all the transverse coordinates of its sources, even though this dependence has not been indicated explicitly in eq.(\ref{singleg}). This integration insures transverse translational invariance of the initial state.

Our goal now is to rewrite this observable in the language of the degrees of freedom of the Reggeon Field Theory. First off, we note that due to color neutrality of the target and the projectile we can write
\begin{eqnarray}
\langle W_T^{ Y-\eta}\vert
\left[(S_z^{A\dagger}\,-\,S_x^{A\dagger})( S^A_{\bar z}\,-\, S^A_y)\right]^{ab}\rangle&=&\frac{1}{N_c^2}\delta^{ab}
\langle W_T^{ Y-\eta}\vert{\rm Tr}\left[(S_z^{A\dagger}\,-\,S_x^{A\dagger})( S^A_{\bar z}\,-\, S^A_y)\right]\rangle\nonumber\\
\langle W_P^\eta\vert J_L^a[x]\,J_R^b[y]\rangle&=&\frac{1}{N_c^2}\delta^{ab}\langle W_P^\eta\vert J_L^c[x]\,J_R^c[y]\rangle
\end{eqnarray}
This leads to
\begin{eqnarray}\label{single}
\frac{d\sigma}{d\eta\,dk^2}&=&\frac{\alpha_s}{ 4\,\pi^3N_c^2}\,
\int_{b;z,\bar z} e^{i\,k(z\,-\,\bar z)}\,\int_{x,y} \frac{(z-x)_i}{(z-x)^2}\,
\frac{(\bar z-y)_i}{(\bar z-y)^2} \,\times\nonumber \\
&\times&\langle W_T^{ Y-\eta}\vert
{\rm Tr}\left[(S_z^{A\dagger}\,-\,S_x^{A\dagger})( S^A_{\bar z}\,-\, S^A_y)\right]\rangle\langle W_P^\eta\vert J_L^a[x]\,J_R^a[y]\rangle
\end{eqnarray}

\begin{figure}
\begin{center}
\centerline{\includegraphics[scale=0.8]{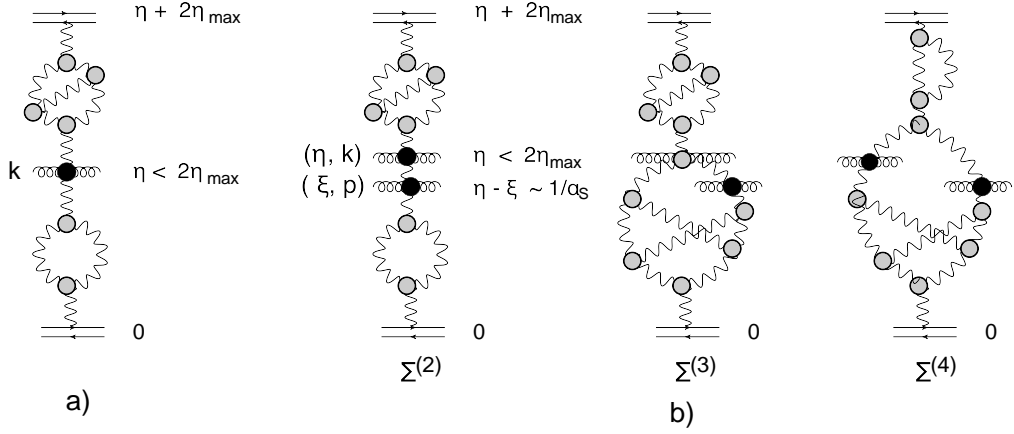}}
\end{center}
\caption{\label{incl}  Mueller diagrams \protect\cite{MUDI} for single ( \fig{incl}-a) and double (\fig{incl}-b) production. The gray circles denote the triple Pomeron vertex while the black blobs describe the emission of the gluon from the BFKL Pomeron.}
\end{figure}

The operators that appear in eq.(\ref{single}) are strictly speaking different than the ones used to define the Reggeon operators discussed in \cite{last}. All the Reggeon and Reggeon conjugate operators discussed in \cite{aklp} are singlets under the $SU_R(N_c)\times SU_L(N_c)$ transformation. On the other hand the operator $J_L^a[x]\,J_R^a[y]$, while a singlet under the vector $SU(N_c)$ subgroup, transforms as an adjoint under the action of the left and right rotations separately. This operator represents the cut Pomeron as discussed in \cite{cutpom}. The reason ref. {\cite{last} discussed only evolution of left and right singlets, is that for the calculation of the forward scattering amplitude of any color singlet physical state, the weight functional has the form
\begin{equation}\label{w}
W\,=\,W[P, O, B...]\delta[\rho]
\end{equation}
where $P, O, B,\ \ etc...$ are all $SU_R(N_c)\times SU_L(N_c)$  singlets. Thus only the evolution of singlet operators is relevant for the calculation of forward scattering amplitude. On the other hand, it is not true that only $SU_R(N_c)\times SU_L(N_c)$ singlets have nonvanishing expectation value in the state specified by $W$.  A trivial example of an operator which is not a singlet, but whose average does not vanish is 
$R^{\alpha\beta}$. Recall that one of the defining properties of $W$ is its normalization \cite{yinyang}
\begin{equation}
\int d\rho \, W\,=\,1\,.
\end{equation}
From this it immediately follows that
\begin{equation}
\int d\rho \,W\, R^{\alpha\beta}(x)\,=\,\delta^{\alpha\beta}
\end{equation}
Thus in this sense, a state specified by any normalized $W$ of the form eq.(\ref{w}) breaks spontaneously the $SU_L(N_c)\times SU_R(N_c)$ symmetry of $H_{RFT}$ down to the diagonal $SU(N_c)$ subgroup \cite{reggeon}.

 The operator $J_L^a[x]\,J_R^a[y]$ therefore has a nonvanishing expectation value even though it is not separately left and right invariant. 
Nevertheless, if at the initial rapidity $W[P, O, B...]$ depends only on $SU_R(N_c)\times SU_L(N_c)$ invariant operators,  it will continue to have this properety at any rapidity, and thus only the evolution of these operators is important. We conclude therefore, that even though our observable is not $SU_L(N_c)\times SU_R(N_c)$ invariant, in terms of evolution it is sufficient to consider only the  invariant Reggeons $P, \ O, etc.$, as was done in \cite{last}. 

The situation may well be different if we were to consider an observable which measures two (or more) gluons separated by a large rapidity interval. For an observable like this one may need to consider explicitly the evolution of $SU_L(N_c)\times SU_R(N_c)$ noninvariant observables which are still invariant under the diagonal $SU_V(N_c)$ between the rapidities of the two observed gluons.
In this paper we do not consider such observables and therefore will not dwell on the evolution of left-right noninvariant operators.

To rewrite eq.(\ref{single}) in terms of the Reggeon operators we first note that on the target side the algebra is straightforward
\begin{eqnarray}\label{ap}
&&\frac{1}{N_c^2}{\rm Tr}\left[(S_z^{A\dagger}\,-\,S_x^{A\dagger})( S^A_{\bar z}\,-\, S^A_y)\right]=-2\bar P(z\bar z)+2\bar P(zy)+2\bar P(x\bar z)-2\bar P(xy)\nonumber\\
&+&\bar P^2(z\bar z)-\bar P^2(zy)-\bar P^2(x\bar z)+\bar P^2(xy) -\bar O^2(z\bar z)+\bar O^2(zy)+\bar O^2(x\bar z)-\bar O^2(xy)\nonumber\\
&\equiv& -\bar P_A(z\bar z)+\bar P_A(zy)+\bar P_A(x\bar z)-\bar P_A(xy)
\end{eqnarray}
where for compactness we have defined "adjoint Pomeron" as $\bar P_A(z\bar z)\equiv 2\bar P(z\bar z)-\bar P^2(z\bar z)+\bar O^2(z\bar z)$.

On the projectile side, it is easy to see that for any $SU_V(N_c)$ invariant function $W_P$,
\begin{equation}\label{jljr}
\langle W_P^\eta\vert J_L^a[x]\,J_R^a[y]\rangle=\langle W_P^\eta\vert J_R^a[x]\,J_R^a[y]\rangle=\langle W_P^\eta\vert J_L^a[x]\,J_L^a[y]\rangle
\end{equation}
Using eq.(\ref{rhorho}) we can therefore write 
\begin{equation}\label{jj1}
 \frac{1}{N_c}J_L^a[x]\,J_R^a[y]\rightarrow P^\dagger_{xy}-\delta_{xy}\int_zP^\dagger_{xz}
 \end{equation}
 
 Substituting eqs.(\ref{ap},\ref{jj1}) into eq.(\ref{single}) we obtain
 \begin{eqnarray}\label{single1}
\frac{d\sigma}{d\eta\,dk^2}&=&\frac{\alpha_s N_c}{4\,\pi^3}\,
\int_{b;z,\bar z} e^{i\,\vec k\cdot(\vec z\,-\,\vec{\bar z})}\,\int_{x,y} \frac{(z-x)_i}{(z-x)^2}\,
\frac{(\bar z-y)_i}{(\bar z-y)^2} \,\times\\
&\times&\Big( -\bar P^T_A(z\bar z)+\bar P^T_A(zy)+\bar P^T_A(x\bar z)-\bar P^T_A(xy)\Big)\Big( P^{\dagger P}_{xy}-\delta_{xy}\int_uP^{\dagger P}_{xu}\Big) \nonumber
\end{eqnarray}
In eq.(\ref{single1}) we have restored the superscripts $T$ and $P$ on the appropriate operators, in order to indicate that they depend on different degrees of freedom. Thus $\bar P^T_A$ denotes an adjoint dual Pomeron operator, which depends on the color charge density of the target $\rho_T$, while $P^{\dagger P}$ denotes conjugate Pomeron which depends on the projectile color charge density $\rho_P$. 
Also we have dropped the reference to the projectile and target wave functions in eq.(\ref{single1}) for simplicity. It should be understood however, that the right hand side of eq.(\ref{single1}) contains matrix elements of the operators over the appropriate wave functions. Thus we use here the simplified notations
 \begin{equation}\label{averag}
\langle W_T^{ Y-\eta}\vert  \bar P^T_A(z\bar z)\rangle\rightarrow \bar P^T_A(z\bar z); \ \ \ \ \ \ \ \ \ \ \ \ \ 
 \langle W_P^\eta\vert P^{\dagger P}_{xy}\rangle\rightarrow P^{\dagger P}_{xy}
 \end{equation}
 and similarly for products of Pomerons and other Reggeons in the remainder of this chapter\footnote{To avoid ambiguity we note that in eq.(\ref{simplif2}) and all other equations in this chapter we encounter a single matrix element of products of projectile operators over the projectile wave function $W_P$, and a single matrix element of products of target operators over the target wave function (as opposed to products of matrix elements).}.
 
 Equation (\ref{single1}) is the correct form of the inclusive single gluon cross section when the gluon is emitted close to the dilute projectile. In this case only the leading order term in weak field expansion contributes to the projectile side matrix element, while the full nonperturbative expression $\bar P^T$ must be kept on the target side, as the target is not necessarily dilute. Conversely, for gluon emitted close to the target, the target field can be expanded, but the full expression must be kept on the projectile side. In that case the observable is given by eq.(\ref{single1}) where  the form of the operators on the target and projectile sides are interchanged. One can in fact generalize eq.(\ref{single1}) to the expression valid for gluon production at any rapidity between the target and the projectile by writing $P^\dagger$ in terms of $\bar P$ {\it a la} eq.(\ref{bardagger}). This procedure in principle is not unique, since there can be different functions of $\bar P$ that upon expansion to leading order reduce to $P^\dagger$. Nevertheless, it makes sense to adopt the simplest expression that reproduces both limits - of dilute projectile and dilute target:
 \begin{equation}
 P^\dagger_{xy}-\delta_{xy}\int_zP^\dagger_{xz}\rightarrow  \frac{2}{g^4}\nabla^2_x\nabla^2_y\bar P_A(x,y)
 \end{equation}
 We can thus write eq.(\ref{single1}) in the form
\begin{eqnarray}\label{single2}
\frac{d\sigma}{d\eta\,dk^2}&=&\frac{N_c}{32\,\pi^5\alpha_s}\,
\int_{b;z,\bar z} e^{i\,\vec k\cdot(\vec z\,-\,\vec{\bar z})}\,\int_{x,y} \frac{(z-x)_i}{(z-x)^2}\,
\frac{(\bar z-y)_i}{(\bar z-y)^2} \,\times\\
&\times&\Big( -\bar P^T_A(z\bar z)+\bar P^T_A(zy)+\bar P^T_A(x\bar z)-\bar P^T_A(xy)\Big)\nabla^2_x\nabla^2_y\bar P_A^P(x,y) \nonumber
\end{eqnarray}
This expression upon some algebra can be cast into an explicitly symmetric form (for details of the derivation see Appendix B)
\begin{equation}\label{single3}
\frac{d\sigma}{d\eta\,dk^2}=\frac{N_c}{8\,\pi^3\alpha_s}\,\frac{1}{k^2}
\int_{b;x,y} e^{i\,\vec k\cdot (\vec x-\vec y)} \frac{\partial}{\partial x_i}\frac{\partial}{\partial y_j}\bar P^T_A(x,y)\Big[ \delta^{ij}\delta^{kl}+\delta^{ik}\delta^{jl}-\delta^{il}\delta^{jk}\Big]\frac{\partial}{\partial x_k}\frac{\partial}{\partial y_l}\bar P_A^P(x,y) 
\end{equation}
 
As mentioned above, integration over the impact parameter $b$ insures translational invariance of the projectile and target distributions. As a result, both averages depend only of the coordinate differences $\vec{x} - \vec{y}$, and eq.(\ref{single3}) can be simplified to
 \begin{equation}\label{single4}
\frac{d\sigma}{d\eta\,dk^2}=\frac{N_c}{8\,\pi^3\alpha_s}\,\frac{1}{k^2}
\int_{x-y} e^{i\,\vec k\cdot(\vec x-\vec y)} \nabla^2 \bar P^T_A(x- y) \nabla^2 \bar P_A^P(x-y) 
\end{equation} 
where now
\beq
P^{T(P)}_A(x- y)\equiv\int_{x+y}P^{T(P)}_A(x,y)
\end{equation}

\subsection{Inclusive two gluon production}
Now let us consider inclusive two gluon production \cite{JK}. Again we first concentrate to the situation when both gluons are produced at rapidity close to that of the projectile. The expression for two gluon production cross section in this limit has been obtained in \cite{Baier,gluons}. It is given by the square of the two gluon production operator amplitude $A(k,p)$ averaged over the projectile and target wave functions
\begin{eqnarray}
A^{ab}_{ij}(k,p)\propto && \int_{u,z}e^{ikz+ipu}\int_{x_1, x_2} 
    \Big\{ {( z -  x_1)_i\over (z-x_1)^2}\,
 \left[S( x_1)-S( z)\right]^{ac}\rho^c(x_1)\Big\} \times \nonumber \\
 && \times \Big\{{(u - x_2)_j\over (u-x_2)^2}\,\left[ S( u)- S(x_2)\right]^{bd}\rho^d(x_2)\Big\}\nonumber\\
&&-{1\over 2} \int_{x_1} 
{(z -  x_1)_i\over (z-x_1)^2}
 {(u-  x_1)_j\over (u-x_1)^2}
\Big\{\left[S( x_1)-S( z)\right]\tilde\rho( x_1)
\left[ S^\dagger(u)+S^\dagger(x_1)\right]\Big\}^{ab}\nonumber\\
&& + \int_{ x_1} 
{( z - u)_i\over (z-u)^2}
{( u - x_1)_j\over (u-x_1)^2}
 \left\{ \left( S( z) - S( u) \right)
 \tilde\rho( x_1) S^{\dagger}( u)\right\}^{ab} \,
\end{eqnarray}
with $\rho$ standing for $\rho_P$ and $\tilde \rho^{ab}=t^c_{ab}\rho^c$.

Squaring the amplitude we get a sum of several terms, which we can separate into three types, according to the power of color charge density
\begin{equation}
\Sigma^{(2)}\propto \rho^2; \ \ \ \ \Sigma^{(3)}\propto \rho^3; \ \ \ \ \ \Sigma^{(4)}\propto \rho^4
\end{equation}
This expression is somewhat schematic. In fact the factors of $\rho$ that belong to the amplitude turn into right color charges $J_R$, while the ones that belong to the conjugate amplitude turn into left color charges $J_L$. To express this in terms of Reggeons we need, just like in the case of single gluon production cross section, relate expectation values of $SU_V(N_c)$ invariant operators to those of $SU_R(N_c)\times SU_L(N_c)$ operators. In the present case it is somewhat more involved than for the single inclusive production. We leave full analysis of the two gluon production amplitude for future work. 
In this paper we concentrate exclusively on the term $\Sigma^{(4)}$, for which the transition to Reggeon operators is straightforward. 
This term is particularly interesting since it gives the leading contribution to correlated gluon emission as discussed in \cite{KLreview}. Approximate numerical implementation of this term has been used to describe the observed ridge correlations in p-p and p-A collisions at LHC in \cite{ridge}.

The contribution of this term to the cross section is
\begin{equation}
\frac{d\sigma^{(4)}}{ d\eta\,dk^2 d\,d\xi\,dp^2}=\left(\frac{\alpha_s}{ 4\,\pi^3}\right)^2\,\Sigma^4(k,p)
\end{equation}
\begin{eqnarray}
\Sigma^{(4)}(k,p)&=&\int_{b;z,\bar z,w,\bar w} e^{i\,k(z\,-\,\bar z)}e^{i\,p(w\,-\,\bar w)}\,\int_{x,y} \frac{(z-x)\cdot(\bar z-y)}{ (z-x)^2(\bar z-y)^2}\,
\frac{(w-u)\cdot(\bar w-v)}{ (w-u)^2(\bar w-v)^2} \,    
\times\nonumber \\
&\times&\langle W_T^{ Y-\eta}\vert
\left[(S_z^{A\dagger}\,-\,S_x^{A\dagger})( S^A_{\bar z}\,-\, S^A_y)\right]^{ab}\, 
\left[(S_w^{A\dagger}\,-\,S_u^{A\dagger})( S^A_{\bar w}\,-\, S^A_v)\right]^{cd}\rangle\ \times \nonumber \\
&\times& \langle W_P^\eta\vert \frac{1}{4}\Big(J_L^a[x]\,J_L^c[u]+J_L^c[u]\,J_L^a[x]\Big)\,\Big(J_R^b[y]\,J_R^d[v]+J_R^d[v]J_R^b[y]\Big)\rangle 
\label{two}
\end{eqnarray}
We express this in terms of the target reggeons in Appendix C. The result for the $\Sigma^{(4)}$ part is 
\begin{eqnarray}\label{simplif2}
\Sigma^{(4)}&=&\left(\frac{1}{32\,\pi^3\alpha_sN_c}\right)^2\,\frac{1}{k^2}\frac{1}{p^2}
\int_{b;x,y,u,v}\cos \vec k\cdot (\vec x-\vec y)\cos \vec p\cdot (\vec u-\vec v) \times\nonumber\\
&\times&\Bigg\{\frac{1}{4}\frac{\partial}{\partial (ij\bar i\bar j)}[\bar P^T_A(x,y)\bar P^T_A(u,v)]\Delta^{ijkl}\Delta^{\bar i\bar j\bar k\bar l}\frac{\partial}{\partial (kl\bar k\bar l)}[\bar P_A^P(x,y) \bar P_A^P(u,v)]\nonumber\\
 &&-\frac{8}{N_c^2}\frac{\partial}{\partial (ij\bar i\bar j)}[\bar d^T_{xy}\bar d^T_{uv}\bar Q^T_{yuvx}]\Delta^{ijkl}\Delta^{\bar i\bar j\bar k\bar l}\frac{\partial}{\partial (kl\bar k\bar l)}[\bar d^P_{yx}\bar d^P_{vu}\bar Q^P_{xvuy}]\Bigg\}
\end{eqnarray}
where 
\begin{eqnarray}
\frac{\partial}{\partial (ijkl)}\equiv \frac{\partial}{\partial x_i}\frac{\partial}{\partial y_j}\frac{\partial}{\partial u_{\bar i}}\frac{\partial}{\partial v_{\bar j}}\, ; \ \ \ \ \ \ \ \ \ \ \ \ \ \ \ \
\Delta^{ijkl}\equiv\delta^{ij}\delta^{kl}+\delta^{ik}\delta^{jl}-\delta^{il}\delta^{jk}
\end{eqnarray}
The second term can now be expressed in terms of dual Pomerons and $B$-Reggeons if desired.

Just like in the case of single gluon production, the integration over the impact parameter insures translational invariance of the projectile and target states. We can thus write
\begin{eqnarray}\label{simplif3}
&&\Sigma^{(4)}=\left(\frac{1}{32\,\pi^3\alpha_sN_c}\right)^2\,\frac{1}{k^2}\frac{1}{p^2}
\int_{x-y,u-v, x+y-u-v} \cos \vec k\cdot (\vec x-\vec y)\cos \vec p\cdot (\vec u-\vec v) \times\\
&&\Bigg\{\frac{1}{4}\frac{\partial}{\partial (ij\bar i\bar j)}D^T_{2A}(x-y,u-v,x+y-u-v)\Delta^{ijkl}\Delta^{\bar i\bar j\bar k\bar l}\frac{\partial}{\partial (kl\bar k\bar l)}D^P_{2A}(x-y,u-v,x+y-u-v)\nonumber\\
 &&-\frac{8}{N_c^2}\frac{\partial}{\partial (ij\bar i\bar j)} D^T_{B}(x-y,u-v,x+y-u-v)\Delta^{ijkl}\Delta^{\bar i\bar j\bar k\bar l}\frac{\partial}{\partial (kl\bar k\bar l)} D^P_{B}(y-x,v-u,x+y-u-v)]\Bigg\}\nonumber
\end{eqnarray}
with
\begin{eqnarray}
&&D_{2A}(x-y,u-v,x+y-u-v)\equiv\int d^2(x+y+u+v)/4\ \ \ \ \bar P_A(x,y)\bar P_A(u,v); \nonumber\\
&&  D_{B}(x-y,u-v,x+y-u-v)\equiv\int d^2(x+y+u+v)/4\ \ \ \ \ \bar d_{xy}\bar d_{uv}\bar Q_{yuvx}
\end{eqnarray}
The functions $D_{2A}(X,Y,Z)$ and $D_B(X,Y,Z)$ in general have nontrivial dependence on all three coordinates, and thus translational invariance does not allow in this case any further simplifications similar to eq.(\ref{single4}).

Superficially it may look like the second term in eq.(\ref{simplif2}) is suppressed in the large $N_c$ limit relative to the first one. However, in the region of applicability of eq.(\ref{simplif2}), where one of the colliding objects is dilute, this is not the case. Consider, for example the dilute projectile limit. As we have discussed above, in this limit
\begin{eqnarray}
&&\nabla^2_x\nabla^2_y\bar P_A(x,y)\rightarrow g^4P^\dagger (x,y)\sim O(\alpha_s^2)\\
&&\nabla^2_x\nabla^2_y \nabla^2_u\nabla^2_v[\bar d_{xy}\bar d_{uv}\bar Q_{yuvx}]\rightarrow g^8 N_c^2 \left[-B^\dagger_{yuvx}+C^\dagger_{yuvx}\right]\sim O(\alpha_s^4N_c^2)\sim O(\alpha_s^2\lambda^2)\nonumber
\end{eqnarray}
Thus both terms in eq.(\ref{simplif2}) are of the same order in the coupling constant and $N_c$. The suppression of the second term comes to fore only in the formal limit where both colliding objects are dense. In this limit, however, expression eq.(\ref{simplif2}) is not valid.

The dense-dense regime is outside the scope of our approximation.  It is possible that although eq.(\ref{simplif2}) is not literally valid in the dense-dense limit, a similar expression in terms of Pomerons and B-Reggeons can be written down with the same $N_c$ counting as in eq.(\ref{simplif2}). In that case the double gluon production in this regime would be dominated by the two Pomeron term. This, however, requires a separate investigation.

\section{Discussion and Conclusions.}
In this paper we clarified the applicability range of the RFT Hamiltonian derived in \cite{aklp} and also showed that in this regime it can be written in a much simpler form in terms of QCD Reggeon degrees of freedom.  We have shown that the approximation of \cite{aklp} is justified when at every step in the evolution at least one of the colliding objects is dilute. The total rapidity range of the evolution is therefore restricted by the nature  of the colliding objects at initial rapidity.  For example, if at initial rapidity $Y_0$  both objects are dilute (dipole-dipole scattering), $H_{RFT}$ can be used to evolve the system up to final rapidity $Y-Y_0=2\eta_{max}$ with $\eta_{max}=\frac{1}{4\ln 2\bar\alpha_s}\ln\frac{1}{\alpha^2_s}$. If on the other hand at $Y_0$ we consider the scattering of a dilute object on a dense one ( dipole-nucleus), the maximal rapidity interval is ony $Y-Y_0=\eta_{max}$. This underscores the known fact that JIMWLK evolution can only be used in a restricted rapidity range of the width $\eta_{max}$.

The limitation of this nature also applies to other existing  approaches, for example to the approach based on classical solution of Yang-Mills equations of motion developed in \cite{raju}. Consider  scattering of two dense objects. The evolution of the scattering matrix can not be calculated in the approach of \cite{raju}. However \cite{raju} showed that one can indeed calculate observables of the type of $n$-gluon inclusive amplitudes, as long as $n\ll 1/\alpha_s$. The reason this is possible is naturally understood in the context of our current discussion. A typical observable has the form\footnote{The "factorizable" form is not essential to our argument. An observable can contain a sum of several factors of the type discussed here. Our argument holds as long as each term in the sum is of the same nature as in eq.(\ref{gvobs}).} similar to eq.(\ref{single})
\beq\label{gvobs}
O^n_{Y,\eta}= K\langle W_T^{ Y-\eta}\vert O_n^T
\rangle\langle W_P^\eta\vert O_n^P\rangle
\eeq
were $K$ is a kinematical factor, while $O^T$ and $O^P$ are some operators constructed from the target and the projectile degrees of freedom. The operators $O_n^T$ and $O_n^P$ are not explicitly given in \cite{raju}, but rather are determined by solving classical equations of motion. However it is clear that for small $n$ these contain a small number of factors of the matrix $S$. In our expression eq. (\ref{single})  this property is explicit, 
\beq
O_1^T={\rm Tr}\left[(S_z^{A\dagger}\,-\,S_x^{A\dagger})( S^A_{\bar z}\,-\, S^A_y)\right]; \ \ \ \ \ \ O_1^P=J_L^a[x]\,J_R^a[y]\,.
\eeq
The averaging of the projectile and target observables decouple from each other in eq.(\ref{gvobs}). Both averages have the form reminiscent of the $S$-matrix of the dense-dilute system. For example $\langle W_T^{ Y-\eta}\vert {\rm Tr}\left[(S_z^{A\,\dagger}\,-\,S_x^{A\,\dagger})( S^A_{\bar z}\,-\, S^A_y)\right] \rangle$ is identical to evaluation of the scattering matrix of a superposition state of an adjoint dipole on the dense target with the wave function $W_T^{Y-\eta}$. The projectile side observables have a very similar structure. 
In the approach of \cite{raju} both, the evolution in $Y-\eta$ and in $\eta$ is taken to be JIMWLK evolution.

As we have discussed above the evolution of such an observable is well approximated by JIMWLK evolution of $W_T$ only as long as the dilute object in the amplitude stays dilute. This means that given initial wave function $W_T^{0}$, one can evolve the one-gluon inclusive amplitude with JIMWLK evolution only up to rapidity $Y-\eta=\eta_{max}$. The same holds for the evolution on the projectile side, limiting the allowed range of $\eta$ to $\eta<\eta_{max}$. 
The maximal rapidity shrinks as the number of gluons in the observable becomes large, since the observable itself becomes "dense". For the $n$-gluon inclusive  production the allowed range is $\eta<\eta_{max}-\frac{1}{4\ln 2\bar\alpha_s}\ln n$ .
Thus, just like for the RFT Hamiltonian of \cite{aklp}, the  applicability of the approach of \cite{raju} is limited to the same range of rapidities.

Interestingly, a similar argument tells us that for certain observables in the case of dilute-dilute scattering our approach is valid in a wider rapidity interval.
Consider an inclusive $n$-gluon production with $n\ll \frac{1}{\alpha^2_s}$, where all gluons are emitted at mid-rapidity $\eta=Y/2$ . This observable is of the same form as eq.(\ref{gvobs}). Here again the averaging  over the projectile and target factorizes, and each average has the form of a dilute-dilute $S$-matrix. Thus we can evolve each average with $H_{RFT}$ as long as $\eta<2\eta_{max}$. We conclude that gluon production at mid-rapidity for dilute-dilute scattering can be calculated using $H_{RFT}$ reliably up to  rapidity $Y=4\eta_{max}$. Of course, it still remains true that if we want to be able to calculate gluon emission at {\it all} intermediate rapidities $0<\eta<Y$, the restriction on $Y$ is that of eq.(\ref{max}). 
 
Although we have only proved the validity of $H_{RFT}$ as given in eq.(\ref{rft}) in the restricted range of rapidities, it is tempting to entertain the possibility that it is valid also beyond $2\eta_{max}$. Nothing in the expression of $H_{RFT}$ itself is indicative of its failure at larger rapidities. Thus retaining the same Hamiltonian (with complete nonlinear commutation
relations between $P$ and $\bar P$ and so on) can be a good model in an extended range of
rapidities. The leading correction that is coming from the "wrong" diagrams, as discussed above, induces the four Pomeron vertex via integrating out the $B$-Reggeon. The main effect of this vertex is to modify the intercept of the two Pomeron state by the amount of order $\lambda/N^2_c$. Further corrections will appear due to the contributions of the Reggeons we have neglected. For example, we expect appearance of the vertex of the type $\bar P\bar P\bar P X$, where $X$ is the Reggeon containing 6 Wilson lines. Integrating out $X$ will produce a $3\rightarrow 3$ Pomeron vertex of order  $\lambda^2/N^4_c$. Thus in this scenario, taking into account the B-Reggeon alone extends the applicability of the theory to rapidities
 $ 2 \tilde{\eta}\,=\,2 \eta_{max} + \Delta \eta$ with $\Delta \eta
\propto N^2_c/\lambda$, and in fact further, since the $3\rightarrow 3$ vertex only becomes important at  $\Delta
\eta \propto N^4_c/\lambda^2$ .  This scenario is plausible albeit unproven in the CGC approach. This could be a fruitful avenue to explore  in seeking understanding of the interrelation of the
CGC approach and Reggeon Field Theory.

We conclude discussion with some comments on the inclusive gluon production.  Our expression for two gluon inclusive production is directly relevant to recent calculations of ridge correlations in \cite{ridge}.
 Ref. \cite{ridge} calculates gluon correlations based exclusively on the first term in eq.(\ref{simplif2}). On the other hand it is very likely that the second term in eq.(\ref{simplif2}) contains nontrivial angular correlations between emitted gluons. This term has no contribution that can be associated with the square of the single gluon emission probability, and as such it describes correlations in their pure form . The nontrivial dependence on the coordinates of the Reggeons most likely leeds to nontrivial angular dependence, and thus angular correlations between emitted gluons. 
As we discussed in the text, formally  this term is sub-leading at large $N_c$ in the dense-dense regime. Although eq.(\ref{simplif2}) is not strictly speaking valid in the dense-dense limit,  one may hope that it at least gives correct large $N_c$ counting. In this case the contribution of $B$-reggeon is sub-leading in this regime. However, when one of the objects is dilute, this term contributes at the same order as the two Pomeron exchange (the first term). In fact, in the dilute-dilute regime it's contribution is formally enhanced by $N_c^2$ relatively to that of the two Pomerons. This is a very interesting effect, as it suggests that in the dilute-dilute limit emission is very strongly correlated due to the presence of this term. This point needs further clarification. However it is certainly true that the relative $N_c$ weight of the two terms is different at different densities. Interestingly, as discussed in \cite{KLreview}, one expects significantly correlated emission to arise from dilute rather than dense regime, where one certainly cannot neglect the effect of the second term ($B$-Reggeon) in eq.(\ref{simplif2}). Even in the dense-dense regime in the calculation of \cite{ridge} the correlated emission given by the first (two Pomeron) term in eq.(\ref{simplif2})  arises only at next to leading order in $1/N_c$. At this order the contribution of the $B$-Reggeon term is equally important and has to be accounted for.

\appendix

\section{Self-Duality of RFT.}

While the interaction term in eq. (\ref{simple}) is by construction self-dual,  it is not that explicit in case of the homogeneous term. Here,   we prove
it using the relations between $\bar P$ and $\bar P^\dagger$
\beq
P^\dagger_{x,y}={4\over g^4}\,\nabla^2_x\nabla^2_y\,\bar P_{x,y}
\eeq
We notice that the kernel $M_{x,y;z}$ can be written as
\beq
M_{x,y;z}= \left\{\nabla^i_z \left[\frac{1}{\nabla^2}(xz)\,-\,\frac{1}{\nabla^2}(yz)\right]\right\}\ \left\{\nabla^i_z \left[\frac{1}{\nabla^2}(xz)\,-\,\frac{1}{\nabla^2}(yz)\right]\right\}\
\eeq
Next we write
\begin{eqnarray}
P_{x,z}+P_{y,z}-P_{x,y}&=& \frac{1}{ 2}\,\int_{u,v} \left[\frac{1}{ \nabla^2}(ux)\frac{1}{\nabla^2}(vz)+\frac{1}{\nabla^2}(uz)\frac{1}{\nabla^2}(vx)+
\frac{1}{\nabla^2}(uy)\frac{1}{ \nabla^2}(vz)+\right. \nonumber \\
&+&\left.\frac{1}{\nabla^2}(uz)\frac{1}{\nabla^2}(vy)-
 \frac{1}{ \nabla^2}(uy)\frac{1}{\nabla^2}(vx)-\frac{1}{\nabla^2}(ux)\frac{1}{\nabla^2}(vy)
\right] \,\nabla^2_u\nabla^2_v\,P_{u,v}\nonumber \\
\end{eqnarray}
Integrating by part once, the homogeneous term in eq. (\ref{simple}) can be written as
\begin{eqnarray}
&&\int_{x,y,z} M_{x,y;z}\,[P_{x,z}+P_{y,z}-P_{x,y}] \,P^\dagger_{x,y}\ =\ -\,\frac{1}{ g^4}\int_{x,y,z,u,v} 
\left\{\nabla^i_z \left[\frac{1}{\nabla^2}(xz)\,-\,\frac{1}{\nabla^2}(yz)\right]^2 \right\}\, \nonumber \\
&&\times\nabla^i_z \, \left[\frac{1}{\nabla^2}(ux)\frac{1}{ \nabla^2}(vz)+\frac{1}{\nabla^2}(uz)\frac{1}{\nabla^2}(vx)+
\frac{1}{ \nabla^2}(uy)\frac{1}{ \nabla^2}(vz)+
\frac{1}{\nabla^2}(uz)\frac{1}{\nabla^2}(vy)\right]\nonumber \\
&&\times\,\nabla^2_u\nabla^2_v\,P_{u,v}\,
\nabla^2_x\nabla^2_y\,\bar P_{x,y}
\end{eqnarray}
Integrating  by parts second time we get
\begin{eqnarray}
&&\int_{x,y,z} M_{x,y;z}\,[P_{x,z}+P_{y,z}-P_{x,y}] \,P^\dagger_{x,y}\ =\ \frac{1}{ g^4}\int_{x,y,z,u,v} 
 \left[\frac{1}{\nabla^2}(xz)\,-\,\frac{1}{\nabla^2}(yz)\right]^2 \, \nonumber \\
&& \left[\frac{1}{ \nabla^2}(ux)\delta_{vz}+\delta_{uz}\frac{1}{\nabla^2}(vx)+
\frac{1}{ \nabla^2}(uy)\delta_{vz}+
\delta_{uz}\frac{1}{\nabla^2}(vy)\right]\,\nabla^2_u\nabla^2_v\,P_{u,v}\,
\nabla^2_x\nabla^2_y\,\bar P_{x,y}
\end{eqnarray}
Integrating over $z$
\begin{eqnarray}
&&\int_{x,y,z} M_{x,y;z}\,[P_{x,z}+P_{y,z}-P_{x,y}] \,P^\dagger_{x,y}\ =\nonumber \\ 
&&~~~~~~~~=\frac{1}{ g^4}\int_{x,y,u,v} 
\left\{ \left[\frac{1}{\nabla^2}(xv)\,-\,\frac{1}{\nabla^2}(yv)\right]^2 \, 
\left[\frac{1}{ \nabla^2}(ux)+\frac{1}{ \nabla^2}(uy)\right]\right. +\nonumber \\
&&~~~~~~~~+\left. \left[\frac{1}{\nabla^2}(xu)\,-\,\frac{1}{\nabla^2}(yu)\right]^2 \, 
\left[\frac{1}{ \nabla^2}(vx)+\frac{1}{ \nabla^2}(vy)\right]
\right\} \nabla^2_u\nabla^2_v\,P_{u,v}\,
\nabla^2_x\nabla^2_y\,\bar P_{x,y}\,=\nonumber \\
&&~~~~~~~~=\frac{1}{ g^4}\int_{x,y,u,v} 
\left\{ \left(\frac{1}{\nabla^2}(xv)\right)^2\frac{1}{ \nabla^2}(uy)+
\left(\frac{1}{\nabla^2}(yv)\right)^2\frac{1}{ \nabla^2}(ux)\right. -\nonumber \\
&&~~~~~~~~\left. -\,2 \frac{1}{\nabla^2}(xv)\frac{1}{\nabla^2}(yv)
\left[\frac{1}{ \nabla^2}(ux)+\frac{1}{ \nabla^2}(uy)\right]+\left(\frac{1}{\nabla^2}(xu)\right)^2\frac{1}{ \nabla^2}(vy)+\right. \nonumber \\
&&~~~~~~~~+\left. \left(\frac{1}{\nabla^2}(yu)\right)^2\frac{1}{ \nabla^2}(vx)
-2 \frac{1}{\nabla^2}(xu)\frac{1}{\nabla^2}(yu)
\left[\frac{1}{ \nabla^2}(vx)+\frac{1}{ \nabla^2}(vy)\right]\
\right\}\times \nonumber \\
&&~~~~~~~~~~~~~~~~~~~~~~~~~~~~~~~~~~~~~~~~~~~~~~~~~~~~~~~~~~~\times \nabla^2_u\nabla^2_v\,P_{u,v}\,
\nabla^2_x\nabla^2_y\,\bar P_{x,y}
\end{eqnarray}
In the last form the expression is explicitly selfdual under exchange of $P$ and $\bar P$.

\section{Single gluon cross section - getting in shape.}
In this appendix we present the algebra which simplifies the form of the single gluon cross section. We start with eq. (\ref{single2}) and rewrite it in terms of the Fourier transforms 
\begin{eqnarray}
&&
\int_{z,\bar z,x,y,u,v} e^{i\,k(z\,-\,\bar z)}\,\int_{x,y} \frac{(z-x)_i}{(z-x)^2}\,
\frac{(\bar z-y)_i}{(\bar z-y)^2} \Big( -\bar P^T_A(z\bar z)+\bar P^T_A(zy)+\bar P^T_A(x\bar z)-\bar P^T_A(xy) \Big) \times \nonumber \\ \nonumber \\
&& \times \nabla^2_x\nabla^2_y\bar P_A^P(x,y)\,=\,\int_{z,\bar z,x,y,u,v}\frac{d^2l}{4\pi^2} \frac{d^2m}{4\pi^2}\frac{d^2s}{4\pi^2}\frac{d^2t}{4\pi^2}\frac{d^2p}{4\pi^2}\frac{d^2q}{4\pi^2}(2\pi)^2P_A^T(p,q)\frac{l_i}{ l^2}\frac{m_i}{ m^2}s^2t^2P_A^P(s,t)\times \nonumber\\ \nonumber\\
&&\times e^{ik(z-\bar z)+il(z-x)+im(\bar z-y)+isx+ity}\Big[-e^{ipz+iq\bar z}+e^{ipz+iqy}+e^{ipx+iq\bar z}-e^{ipx+iqy}\Big]
\end{eqnarray}
After straightforward integration over the four coordinate variables  and momenta $l,m, p$ and $q$, which realizes the momentum delta functions, we arrive at
\begin{eqnarray}
&&\int \frac{d^2s}{4\pi^2}\frac{d^2t}{4\pi^2} 4\pi^2P_A^T(-k-s, k-t)P_A^P(s,t)s^2t^2\Big(-\frac{s_i}{s^2}-\frac{k_i}{k^2}\Big)\Big(\frac{t_i}{t^2}-\frac{k_i}{k^2}\Big)\nonumber\\
&=&\int \frac{d^2s}{4\pi^2}\frac{d^2t}{4\pi^2} 4\pi^2P_A^T(-k-s, k-t)P_A^P(s,t)\frac{1}{k^2}\nonumber\\
&\times&\Big[-(-k-s)\cdot t(k-t)\cdot s+(-k-s)\cdot(k-t)s\cdot t+(-k-s)\cdot s(k-t)\cdot t\Big]
\end{eqnarray}
Finally, expressing this back through the functions in coordinate representation, and collecting the omitted prefactor we obtain
\begin{equation}\label{simplif}
\frac{d\sigma}{d\eta\,dk^2}=\frac{N_c}{8\,\pi^3\alpha_s}\,\frac{1}{k^2}
\int_{x,y} e^{i\,k(x-y)} \frac{\partial}{\partial x_i}\frac{\partial}{\partial y_j}\bar P^T_A(x,y)\Big[ \delta^{ij}\delta^{kl}+\delta^{ik}\delta^{jl}-\delta^{il}\delta^{jk}\Big]\frac{\partial}{\partial x_k}\frac{\partial}{\partial y_l}\bar P_A^P(x,y) 
\end{equation}

\section{Projecting two-gluon emission operator  onto Reggeons}

In this Appendix we express the two gluon inclusive production cross section eq. (\ref{two}) in terms of Reggeon operators.
Since the projectile and the target are both color neutral, 
we have to project 
\beq
\left[(S_z^{A\,\dagger}\,-\,S_x^{A\,\dagger})( S^A_{\bar z}\,-\, S^A_y)\right]^{ab}\, 
\left[(S_w^{A\,\dagger}\,-\,S_u^{A\,\dagger})( S^A_{\bar w}\,-\, S^A_v)\right]^{cd}
\eeq
and
\beq J_L^a[x]\,J_R^b[y]\,J_L^c[u]\,J_R^d[v] \eeq
onto color singlets separately. A color decomposition of the above expressions using  complete set of projectors has been introduced in ref. \cite{KLreview}.
Here we follow an alternative route \cite{last} which is better adopted to the large $N_c$ limit.  Define
\begin{equation}
t^{abcd}\equiv {\rm Tr} (T^aT^bT^cT^d)
\end{equation}
To leading order in $1/N_c$ one has the following decomposition \cite{last}
\begin{eqnarray}\label{decomp}
&& 
\left[S_x^{A\,\dagger}\, S^A_y\right]^{ab}\, 
\left[S_u^{A\,\dagger}\, S^A_v\right]^{cd}
=\frac{1}{N_c^4}\Bigg[\delta^{ab}\delta^{cd}\Big[D_{2\,traces}(x,y,u,v)\nonumber \\
&&~~~~~~~~~~~~~~~~~~~~~~~~~~~~~-\frac{4}{N_c}[D^{1234}_B(xyuv)+ D^{1432}_B(xyuv)+D^{1243}_B(xyuv)+ D^{1342}_B(xyuv)]\Big]\nonumber\\
&&+\delta^{a c}\delta^{b d}\Big[ D_{1\,trace}(xyvu)-\frac{4}{N_c}[ D^{1432}_B(xyuv)+ D^{1234}_B(xyuv)+ D^{1423}_B(xyuv)+ D^{1324}_B(xyuv)]\Big]\nonumber\\
&&+\delta^{ad}\delta^{bc}\Big[D_{1\,trace}(xyuv)-\frac{4}{N_c}[ D^{1324}_B(xyuv)+ D^{1423}_B(xyuv)+ D^{1342}_B(xyuv)+ D^{1243}_B(xyuv)]\Big]\Bigg]\nonumber\\
&&+\frac{16}{N_c^4}\Bigg[t^{dcba} D^{1234}_B(xyuv)+t^{cdba}D^{1243}_B(xyuv)+t^{dbca} D^{1324}_B(xyuv)\nonumber\\
&&+t^{bdca} D^{1342}_B(xyuv)+t^{cbda} D^{1423}_B(xyuv)+t^{bcda} D^{1432}_B(xyuv)\Bigg]
\end{eqnarray}
\beq
D_{2\,traces}(x,y,u,v)\,=\,
tr[{S_x^{A}}^\dagger\, S^A_y]\, tr[{S_u^{A}}^\dagger\, S^A_v]
\eeq
\beq
D_{1\,trace}(x,y, u,v)\,=\,
tr[{S_x^{A}}^\dagger\, S^A_y\,{S_u^{A}}^\dagger\, S^A_v]
\eeq
\beq
D_{B}^{ijkl}(x,y, u,v)\,=\,
t^{a_ia_ja_ka_l}[{S_x^{A}}^\dagger\, S^A_y]^{a_1a_2}\, [{S_u^{A}}^\dagger\, S^A_v]^{a_3a_4}
\eeq
We can rewrite the above expressions in terms of fundamental dipoles and quadrupoles
\beq
D_{2\,traces}(x,y,u,v)\,=\,(N^2_c\, \bar d_{xy}\,\bar d_{yx}\,-1)\,(N^2_c\, \bar d_{uv}\,\bar d_{vu}\,-1)
\eeq
\beq
D_{1\,trace}(x,y, u,v)\,=\,N^2_c\,\bar Q_{xvuy}\,\bar Q_{yuvx}\,-\,1\,;
\eeq
\beq
D_{1\,trace}(x,y,v, u)\,=\,N^2_c\,\bar Q_{xuvy}\,\bar Q_{yvux}\,-\,1
\eeq
\beq
D_{B}^{1234}(xyuv)\,=\,\frac{N^3_c}{ 4} \left(\bar d_{xy}\,\bar d_{uv}\,\bar Q_{yuvx}\,-\,\frac{1}{ N^2_c} \bar d_{xy}\,\bar d_{yx}\,-\,\frac{1}{ N^2_c} \bar d_{uv}\,\bar d_{vu}\,
+\frac{1}{ N_c^4}\right)
\eeq
\beq
D_{B}^{1432}(xyuv)\,=\,\frac{N^3_c}{4} \left(\bar d_{yx}\,\bar d_{vu}\,\bar Q_{xvuy}\,-\,\frac{1}{ N^2_c} \bar d_{xy}\,\bar d_{yx}\,-\,\frac{1}{ N^2_c} \bar d_{uv}\,\bar d_{vu}\,
+\frac{1}{ N_c^4}\right)
\eeq
\beq
D_{B}^{1342}(xyuv)\,=\,\frac{N^3_c}{ 4} \left(\bar d_{yx}\,\bar d_{uv}\,\bar Q_{xuvy}\,-\,\frac{1}{ N^2_c} \bar d_{xy}\,\bar d_{yx}\,-\,\frac{1}{ N^2_c} \bar d_{uv}\,\bar d_{vu}\,
+\frac{1}{ N_c^4}\right)
\eeq
\beq
D_{B}^{1243}(xyuv)\,=\,\frac{N^3_c}{ 4} \left(\bar d_{xy}\,\bar d_{vu}\,\bar Q_{yvux}\,-\,\frac{1}{ N^2_c} \bar d_{xy}\,\bar d_{yx}\,-\,\frac{1}{ N^2_c} \bar d_{uv}\,\bar d_{vu}\,
+\frac{1}{ N_c^4}\right)
\eeq
\beq
D_{B}^{1324}(xyuv)\,=\,\frac{N_c}{ 4} \left(\frac{1}{ N_c}\, tr[S_u^\dagger S_vS_x^\dagger S_yS_v^\dagger S_uS_y^\dagger S_x]\,-\,
 \bar d_{xy}\,\bar d_{yx}\,-\, \bar d_{uv}\,\bar d_{vu}\,
+\frac{1}{ N_c^2}\right)
\eeq
\beq
D_{B}^{1423}(xyuv)\,=\,\frac{N_c}{ 4} \left(\frac{1}{ N_c}\, tr[S_v^\dagger S_uS_x^\dagger S_yS_u^\dagger S_vS_y^\dagger S_x]\,-\,
 \bar d_{xy}\,\bar d_{yx}\,-\, \bar d_{uv}\,\bar d_{vu}\,
+\frac{1}{ N_c^2}\right)
\eeq

Convoluting $\left[{S_x^{A}}^\dagger\, S^A_y\right]^{ab}\, 
\left[{S_u^{A}}^\dagger\, S^A_v\right]^{cd}$ 
with four $J$s and retaining terms leading in $N_c$ only, we obtain
\begin{eqnarray}\label{lastwo}
&& 
\left[{S_x^{A}}^\dagger\, S^A_y\right]^{ab}\, 
\left[{S_u^{A}}^\dagger\, S^A_v\right]^{cd} \ J^a_1\,J_2^b\,J_3^c\,J_4^d
= (1-\bar P^{T}_{A\,xy})(1-\bar P^{T}_{A\,uv})\ J_1^a\,J_2^a\ J_3^b\,J^b_4\,+ \nonumber\\
&&+\,
\frac{4}{  N_c} \Big[\bar d^T_{xy}\,\bar d^T_{uv}\,\bar Q^T_{yuvx}\, tr[T^dJ_4^d\, T^c J_3^c\,T^bJ_2^b\,T^aJ_1^a]\,+
\bar d^T_{yx}\,\bar d^T_{vu}\,\bar Q^T_{xvuy}\, tr[ T^bJ_2^b\,T^c J_3^c\,T^dJ_4^d\,T^aJ_1^a]\ +
\nonumber \\
&&+ \bar d^T_{yx}\,\bar d^T_{uv}\,\bar Q^T_{xuvy}\, tr[T^bJ_2^b\,T^dJ_4^d\, T^c J_3^c\,T^aJ_1^a]\,+
\bar d^T_{xy}\,\bar d^T_{vu}\,\bar Q^T_{yvux}\, tr[ T^c J_3^c\,T^dJ_4^d\,T^bJ_2^b\,T^aJ_1^a]\Big]
\label{restwo}
\end{eqnarray}
Just like in the single gluon production case, eq.(\ref{jljr}), there is no difference between the expectation values of $J_LJ_LJ_RJ_R$ and $J_LJ_LJ_LJ_L$. We can thus directly express the product of $J$'s in terms of the Reggeons.
As before, we only need to identify this relation in leading order. 
\beq
J_1^a\,J_2^a\ J_3^b\,J^b_4\ \rightarrow\ \frac{4\,N_c^2}{ g^8}\ \nabla^2_1\nabla^2_2\nabla^2_3\nabla^2_4\left[\bar P^{P}_{A\,12}\, \bar P^{P}_{A\,34}\right]
\eeq
\beq\label{4j}
\frac{1}{ N_c}\, tr[T^dJ_4^d\, T^c J_3^c\,T^bJ_2^b\,T^aJ_1^a]\ \rightarrow\frac{1}{g^8}  \nabla^2_1\nabla^2_2\nabla^2_3\nabla^2_4[\bar Q^P_{1432}]\,
\eeq
We will follow the same line as for the single gluon emission, namely, we will write the final expressions in the form which are valid also in the limit of dense projectile and dilute target, in other words in the form which is symmetric under the interchange of $T$ and $P$ labels. In order to do that we modify the previous relation to
\beq\label{4j1}
\frac{1}{ N_c}\, tr[T^dJ_4^d\, T^c J_3^c\,T^bJ_2^b\,T^aJ_1^a]\ \rightarrow\frac{1}{g^8}  \nabla^2_1\nabla^2_2\nabla^2_3\nabla^2_4[\bar d^P_{21}\bar d^P_{43}\bar Q^P_{1432}]\,
\eeq
To leading order in $J$ and $1/N_c$, the right hand side of eq.(\ref{4j1}) coincides with that of eq.(\ref{4j})
Thus 
\begin{eqnarray}\label{lastwo1}
&& 
\left[{S_x^{A}}^\dagger\, S^A_y\right]^{ab}\, 
\left[{S_u^{A}}^\dagger\, S^A_v\right]^{cd} \ J^a_1\,J_2^b\,J_3^c\,J_4^d
=\frac{N_c^2}{ g^8}\Bigg[ (1-\bar P^{T}_{A\,xy})(1-\bar P^{T}_{A\,uv})\nabla^2_1\nabla^2_2\nabla^2_3\nabla^2_4\left[\bar P^{P}_{A\,12}\, \bar P^{P}_{A\,34}\right]
\nonumber \\
&&+\,
\frac{8}{N_c^2}\Big\{\bar d^T_{xy}\,\bar d^T_{uv}\,\bar Q^T_{yuvx}\, \nabla^2_1\nabla^2_2\nabla^2_3\nabla^2_4\left[\bar d^P_{21}\bar d^P_{43}\bar Q^P_{1432}\right]\,+
\bar d^T_{yx}\,\bar d^T_{vu}\,\bar Q^T_{xvuy}\,\nabla^2_1\nabla^2_2\nabla^2_3\nabla^2_4\left[\bar d^P_{41}\bar d^P_{23}\bar Q^P_{1234}\right] 
\nonumber \\
&&+ \bar d^T_{yx}\,\bar d^T_{uv}\,\bar Q^T_{xuvy}\,\nabla^2_1\nabla^2_2\nabla^2_3\nabla^2_4\left[\bar d^P_{31}\bar d^P_{24}\bar Q^P_{1243}\right] \,+
d^T_{xy}\,d^T_{vu}\,Q^T_{yvux}\,\nabla^2_1\nabla^2_2\nabla^2_3\nabla^2_4\left[\bar d^P_{21}\bar d^P_{34}\bar Q^P_{1342}\right]\Big\}\Bigg]\nonumber\\
\end{eqnarray} 
From here the Fourier transform algebra is identical to that for the single gluon and can be done separately for the two sets of momenta corresponding to the two gluons. The four terms  can be easily transformed into each other by change of dummy integration variables. The result is to turn each exponential factor $e^{ikx}$ in the expression anogous to eq.(\ref{single3}) into $2\cos kx$. The resulting expression is 
\begin{eqnarray}\label{simplif1}
\frac{d\sigma}{ d\eta\,dk^2 d\,d\xi\,dp^2}&=&\left(\frac{N_c}{32\,\pi^3\alpha_s}\right)^2\,\frac{1}{k^2}\frac{1}{p^2}
\int_{x,y,u,v} \cos k(x-y)\cos p(u-v) \times \nonumber \\
&\times&\Bigg\{\frac{1}{4}\frac{\partial}{\partial (ij\bar i\bar j)}[\bar P^T_A(x,y)\bar P^T_A(u,v)]\Delta^{ijkl}\Delta^{\bar i\bar j\bar k\bar l}\frac{\partial}{\partial (kl\bar k\bar l)}[\bar P_A^P(x,y) \bar P_A^P(u,v)]\nonumber\\
 &&-\frac{8}{N_c^2}\frac{\partial}{\partial (ij\bar i\bar j)}[\bar d^T_{xy}\bar d^T_{uv}\bar Q^T_{yuvx}]\Delta^{ijkl}\Delta^{\bar i\bar j\bar k\bar l}\frac{\partial}{\partial (kl\bar k\bar l)}[\bar d^P_{yx}\bar d^P_{vu}\bar Q^P_{xvuy}]\Bigg\}
\end{eqnarray}
where we have defined
\begin{eqnarray}
\frac{\partial}{\partial (ijkl)}\equiv \frac{\partial}{\partial x_i}\frac{\partial}{\partial y_j}\frac{\partial}{\partial u_{k}}\frac{\partial}{\partial v_{l}}\,; \  \ \ \ \ \ \ \ \ \ \ \ \ \ \ \ \ 
\Delta^{ijkl}\equiv\delta^{ij}\delta^{kl}+\delta^{ik}\delta^{jl}-\delta^{il}\delta^{jk}
\end{eqnarray}

\section*{Acknowledgments}
T.A. and M.L. thank the Physics Department of the University of Connecticut for hospitality during the visits while the work 
on this project was in progress.  A.K. and T.A. thank  the  Physics Departments of the Ben-Gurion University of the Negev; A.K. also thanks 
  Universidad T\'ecnica Federico Santa Mar\'\i a.
The research  was supported by the DOE grant DE-FG02-13ER41989; the BSF grant 2012124,    Marie Curie Grant  PIRG-GA-2009-256313; the  ISRAELI SCIENCE FOUNDATION grant \#87277111;  the People Programme (Marie Curie Actions) of the European Union's Seventh Framework Programme FP7/2007-2013/ under REA
grant agreement n318921;  the  Fondecyt (Chile) grants  1100648 and 1130549; European Research Council grant HotLHC ERC-2001-
StG-279579; Ministerio de Ciencia e Innovac\'\i on of Spain grants FPA2009-06867-E and Consolider-Ingenio 2010 CPAN CSD2007-00042 and by FEDER.

 \end{document}